\RequirePackage{fix-cm}
\RequirePackage{amsmath}
\documentclass[smallextended]{svjour3}       
\smartqed  
\usepackage[english]{babel}
\usepackage[utf8x]{inputenc}
\usepackage[T1]{fontenc}
\usepackage{eqnarray} 
\usepackage{graphicx}
\usepackage{array, tabularx, caption, boldline}
\usepackage{cellspace}
\usepackage{float}
\usepackage{epstopdf}
\usepackage{color}                    
\usepackage{algorithm}
\usepackage{algpseudocode}            
\usepackage{algorithmicx}
\usepackage{multirow}
\usepackage{multicol}
\usepackage{cite}
\usepackage{lipsum}
\usepackage{caption}
\usepackage{subcaption}
\errorcontextlines\maxdimen

\makeatletter
    \newcommand*{\algrule}[1][\algorithmicindent]{\makebox[#1][l]{\hspace*{.5em}\thealgruleextra\vrule height \thealgruleheight depth \thealgruledepth}}%
\newcommand*{\thealgruleextra}{}
\newcommand*{\thealgruleheight}{.75\baselineskip}
\newcommand*{\thealgruledepth}{.25\baselineskip}

\newcount\ALG@printindent@tempcnta
\def\ALG@printindent{%
    \ifnum \theALG@nested>0
        \ifx\ALG@text\ALG@x@notext
        \else
            \unskip
            \addvspace{-1pt}
            \ALG@printindent@tempcnta=1
            \loop
                \algrule[\csname ALG@ind@\the\ALG@printindent@tempcnta\endcsname]%
                \advance \ALG@printindent@tempcnta 1
            \ifnum \ALG@printindent@tempcnta<\numexpr\theALG@nested+1\relax
            \repeat
        \fi
    \fi
    }%
\usepackage{etoolbox}
\patchcmd{\ALG@doentity}{\noindent\hskip\ALG@tlm}{\ALG@printindent}{}{\errmessage{failed to patch}}
\makeatother

\newbox\statebox
\newcommand{\myState}[1]{%
    \setbox\statebox=\vbox{#1}%
    \edef\thealgruleheight{\dimexpr \the\ht\statebox+1pt\relax}%
    \edef\thealgruledepth{\dimexpr \the\dp\statebox+1pt\relax}%
    \ifdim\thealgruleheight<.75\baselineskip
        \def\thealgruleheight{\dimexpr .75\baselineskip+1pt\relax}%
    \fi
    \ifdim\thealgruledepth<.25\baselineskip
        \def\thealgruledepth{\dimexpr .25\baselineskip+1pt\relax}%
    \fi
    \State #1%
    \def\thealgruleheight{\dimexpr .75\baselineskip+1pt\relax}%
    \def\thealgruledepth{\dimexpr .25\baselineskip+1pt\relax}%
}

\begin{document}

\title{Joint QoS-control and Handover Optimization in Backhaul aware SDN-based LTE Networks}

\author{Furqan Hameed Khan         \and
        Marius Portmann 
}


\institute{Furqan Hameed Khan \at
              School of ITEE, The University of Queensland, Australia \\
              Tel.: -\\
              Fax: -\\
              \email{furqanhameed.khan@uq.edu.au}           
           \and
           Marius Portmann \at
              marius@ieee.org
}

\date{Received: date / Accepted: date}

\maketitle

\begin{abstract}
Future cellular networks will be dense and require key traffic management technologies for fine-grained network control. The problem gets more complicated in the presence of different network segments with bottleneck links limiting the desired quality of service (QoS) delivery to the last mile user. In this work, we first design a framework for software-defined cellular networks (SDCN) and then propose new mechanisms for management of QoS and non-QoS users traffic considering both access and backhaul networks, jointly. The overall SDN-LTE system and related approaches are developed and tested using network simulator (ns-3) in different network environments. Especially, when the users are non-uniformly distributed, the results shows that compared to other approaches, the proposed load distribution algorithm enables at least 6\% and 23\% increase in the average QoS user downlink (DL) throughput for all network users and the aggregate throughput of 40\% users with lowest throughput (edge users), respectively. Also, the proposed system efficiently achieves desired QoS and handles the network congestion without incurring significant overhead.
\keywords{LTE \and Access and Backhaul Networks \and Quality of Service \and Downlink}
\end{abstract}

\section{Introduction}
\label{intro}
Future cellular networks integrate key technologies like heterogeneous networks (HetNets) and massive-MIMO that increase the network density and make user-oriented traffic management more crucial. Together with that, the ever changing loads of neighbouring base stations (BSs) and limited resource availability in access and backhaul network restrict the resulting gains of these technologies. To tackle these challenges, most of the previous works proposes solutions considering interference-limited access networks with infinite backhaul capacity. However, due to increasing data rate demand in future cellular networks, limited backhaul networks can become a bottleneck and this has recently emerged as an important issue~\cite{reference001,reference002,reference003}. A key idea is to consider programmable backhaul and access networks that let network operators program cellular networks to dynamically modify and implement services for user traffic flow management and routing procedures. 

\begin{figure}[t!]
\begin{center}
\resizebox{3.5 in}{!}{\includegraphics{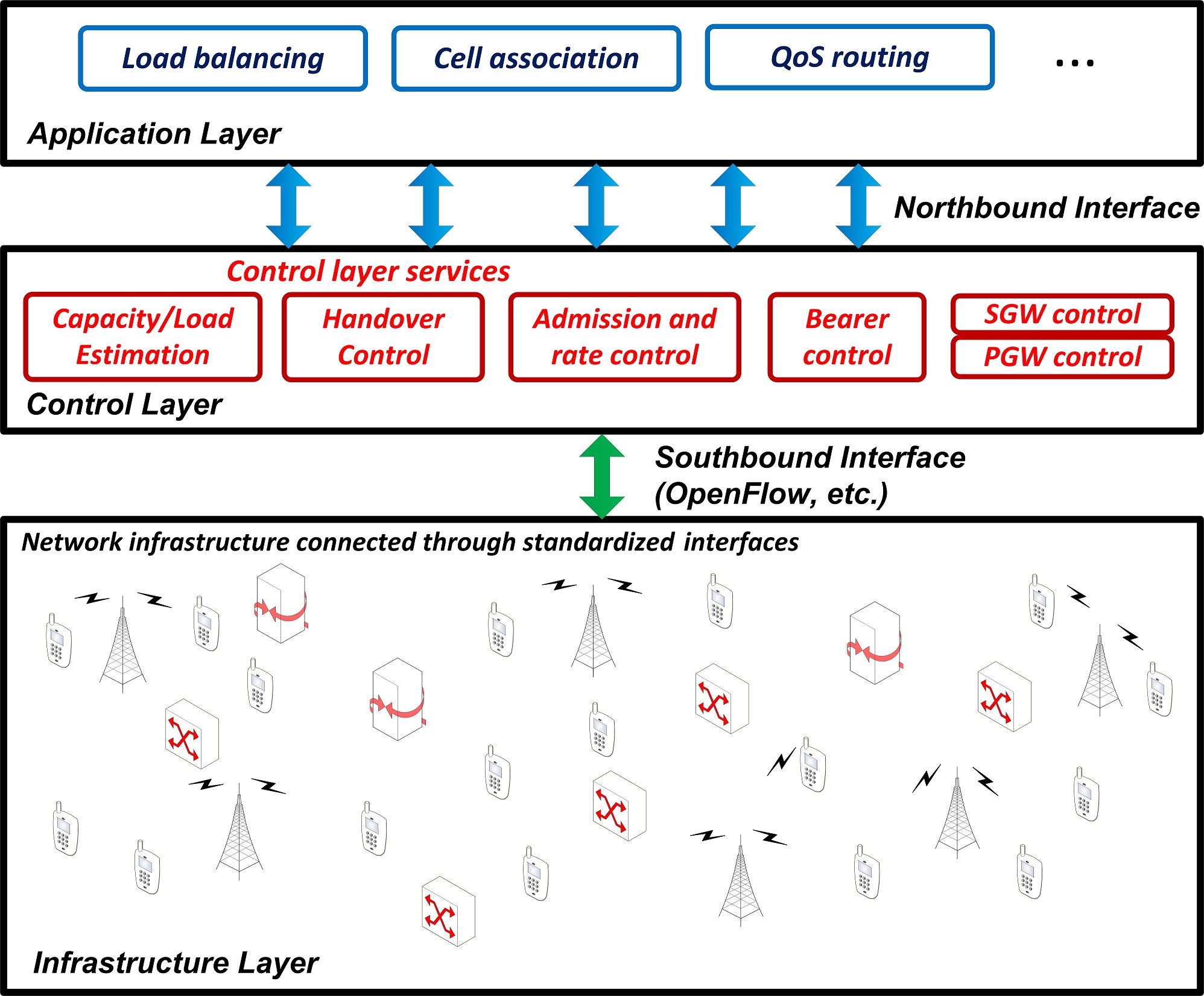}}
\caption{\em Framework for SDN based cellular network~\cite{reference004}. }
\label{jpg1}
\end{center}
\end{figure}

Software-defined networks (SDN) is a new concept that simplify network architecture by separating and centralizing the control functionality from the data forwarding path of networking devices. The SDN inspired cellular network framework in Fig.~\ref{jpg1} shows three layers, namely infrastructure, control, and application. The infrastructure layer consists of user devices, BSs, network switches, and gateways to provide connectivity to the external network. Different control layer services are running over the network infrastructure inside the control layer that enables its programmability through some standardized south-bound interface. These services are connected through some north-bound interface to the application layer running user-defined applications. SDN until now has been mainly applied for cellular networks to reduce the signaling load~\cite{reference005,reference006} and is widely gaining significance for resource management in interference and capacity limited access and backhaul networks~\cite{reference007}. Meanwhile, as part of the efforts of SDN integration in cellular networks integration, the Open Networking Foundation (ONF) is designing frameworks to modify the standard south-bound interface (e.g. OpenFlow) to support mobile networks~\cite{reference008}. Together with that, various recent works have used SDN concepts to improve traffic handling, mobility management, and signaling overhead reduction in cellular networks~\cite{reference005,reference009}.

\begin{figure}[t!]
\begin{center}
\resizebox{3.5 in}{!}{\includegraphics{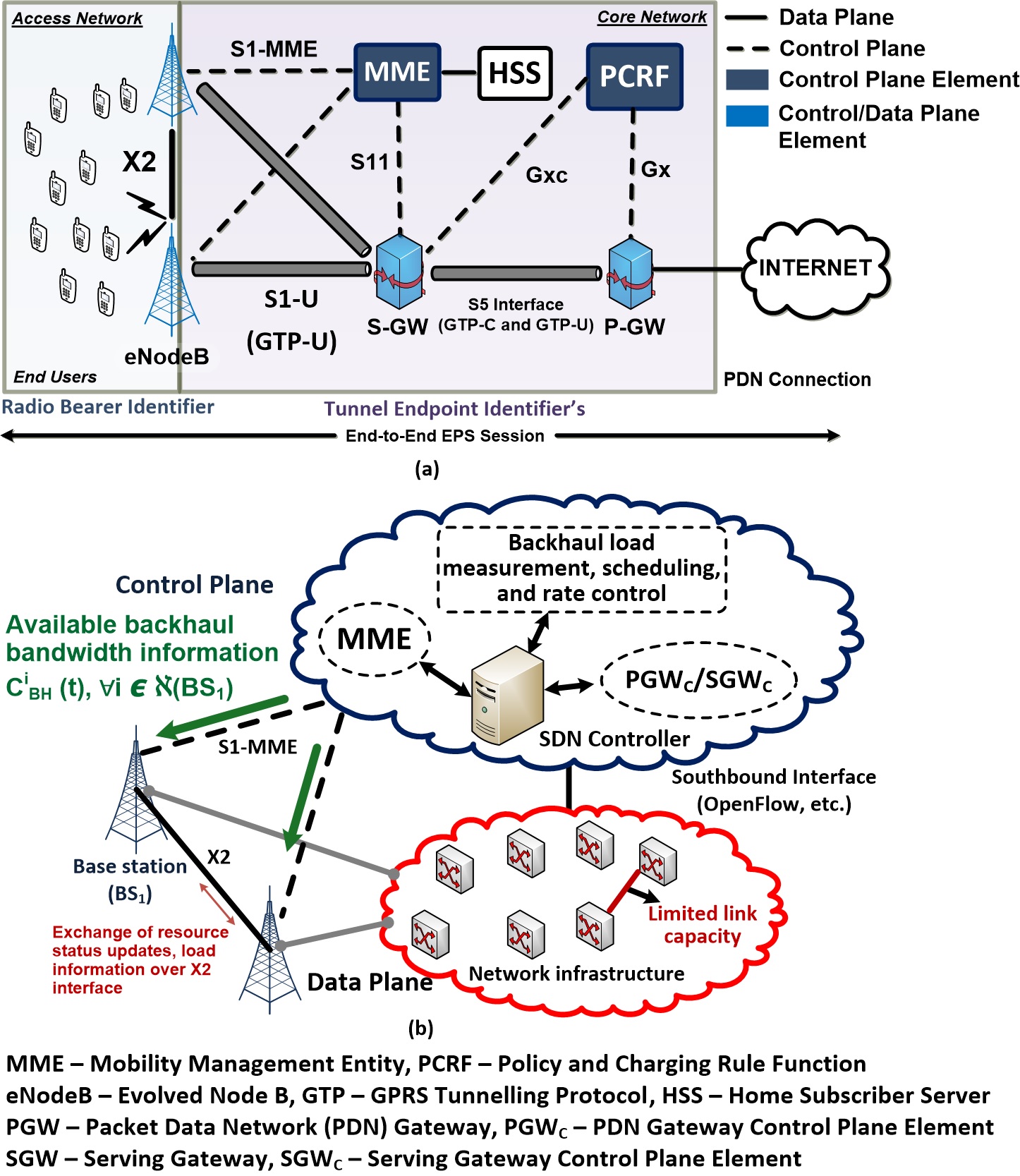}}
\caption{\em LTE network architecture (a) traditional LTE network. (b) software-defined LTE network~\cite{reference004}.}
\label{jpg2}
\end{center}
\end{figure}

Long Term Evolution (LTE) is a broadband wireless access technology that enables an all-ip based end-to-end connectivity for end users. The Fig.~\ref{jpg2}a shows the complete LTE network architecture consisting of access network and the core network. In current LTE networks, the handover decision is made at the specific serving BS, that has local information about its load and user's instantaneous channel condition. From Fig.~\ref{jpg2}a it can be seen that the control and data plane functionalities in LTE are tightly coupled over different network entities. As a consequence, a slight change in access network segment results in immense signaling messages between mobility management entity (MME) and serving gateway (SGW) as well as between SGW and PDN-GW~\cite{reference005}. Note that both cellular access and backhaul network segments have different characteristics and limitations. In the access network, the load of new users, limited bandwidth, and the interference from other cells limits the spectral efficiency, while in backhaul networks, the congestion due to the number of competing flows sharing the available link, limits the resulting performance. This work proposes a SDN based load distribution solution that uses both access and transport backhaul load information for the network-wide performance enhancement. Getting the benefit of SDN, we suggest that the global view of the SDN controller will enable better resource management during critical network operations like handover. In the resulting SDN-enabled LTE architecture as of Fig.~\ref{jpg2}b, the controller monitor user flows in the backhaul network and notify each BS about its available backhaul bottleneck link bandwidth and the bandwidth of the neighbouring BSs. Using such backhaul path information, the BS will make efficient handover decision by avoiding end user throughput degradation due to network congestion.

The following section gives an overview of related works, the research gaps, and the motivation behind programmable cellular networks. Section III presents our system model with the optimization problem, and the relevant assumptions. The details of our proposed load balancing (LB) algorithm are explained in Section IV, while in Section V, we discuss control functionalities in the form of representative modules that we have added to our controller for the backhaul traffic management. Section VI gives the details of experimental setup in Network Simulator (ns-3) with the considered scenarios. Section VII describes the results while Section VIII gives insights on the handover issues in real SDN-LTE networks.
\section{Related Works}
\label{sec:1}
We have divided earlier works into different categories as shown in Table~\ref{tab:1}. First we briefly discuss works considering backhaul and access networks separately, then we summarize the works that consider both access and backhaul jointly.
\begin{table*}[t]
\caption{Contributions using SDN for resource management in cellular networks}\centering
\label{tab:1}
\begin{center}
\resizebox{\textwidth}{!}{%
\begin{tabular}{|c|c|c|c|c|c|}
 \hline
 \textbf{Network Segment} & \textbf{Central Idea} & \textbf{Strong Aspect} & \textbf{Limitations} & \textbf{SDN enabled} & \textbf{QoS/NonQoS/both} \\
 \hline
 \multirow{7}{*}{Access Network} & \begin{tabular}{@{}c@{}}Multi-resource optimization \cite{reference013} \end{tabular} & \begin{tabular}{@{}c@{}}Consider rate, throughput, \\ and delay \end{tabular} & Limited to theoretical modeling & No & No \\
 & QoS-aware \cite{reference011} & \begin{tabular}{@{}c@{}}Reduce blocking probability, \\ achieve load balancing\end{tabular} & Increases average cell load & No & Both \\
 & QoS and channel-aware \cite{reference012}, \cite{reference010} & \begin{tabular}{@{}c@{}}Better load balancing \\ with load reduction \end{tabular} & No system-level procedure & No & Both, Only QoS \\
 & Delay-aware \cite{reference017} & Efficient load balancing & Does not guarantee QoS & Yes & QoS \\
 & Channel-aware \cite{reference014} & \begin{tabular}{@{}c@{}}20-100 \% throughput\\ improvement \end{tabular} & \begin{tabular}{@{}c@{}}Specific system-level \\ details missing \end{tabular} & Limited & Not stated \\
 & \begin{tabular}{@{}c@{}}New load balancing \\mechanism \cite{reference016} \end{tabular} & \begin{tabular}{@{}c@{}}Incentivizing users \\ to change location \end{tabular} & \begin{tabular}{@{}c@{}} User-oriented solution limits \\ the resulting performance \end{tabular} & Yes & No \\
& \begin{tabular}{@{}c@{}} Handover parameters adjustment\\ for conflict avoidance (HPACA)~\cite{reference027}\end{tabular} & \begin{tabular}{@{}c@{}} Reduces call dropping and\\ ping-pong handover rate\end{tabular} & \begin{tabular}{@{}c@{}} Increases signaling overhead\end{tabular} & No & No \\
 \hline
 \multirow{4}{*}{Backhaul Network} & Congestion-aware \cite{reference002}& \begin{tabular}{@{}c@{}}Congestion detection and \\ congestion control \end{tabular} & \begin{tabular}{@{}c@{}}No access network, \\ high overhead \end{tabular} & No & Yes \\
 & QoS-aware \cite{reference009} & QoS in LTE backhaul & No access network & Yes & Yes \\
 & \begin{tabular}{@{}c@{}} Load balancing strategies \cite{reference001} \end{tabular} & \begin{tabular}{@{}c@{}}Efficient global and local \\ load estimation \end{tabular} & \begin{tabular}{@{}c@{}} System-level implementation missing \\ (only numerical analysis) \end{tabular} & No & NonQoS \\
 & \begin{tabular}{@{}c@{}}Backhaul network \\congestion \cite{reference003}\end{tabular} & \begin{tabular}{@{}c@{}}NGBR traffic \\  distribution \end{tabular} & \begin{tabular}{@{}c@{}} No access network \\ consideration \end{tabular} & No & Only nonQoS \\
 \hline
 \multirow{3}{*}{\begin{tabular}{@{}c@{}} Both Backhaul and\\Access Networks \end{tabular}} & Mobility-driven \cite{reference018} & Overload detection & \begin{tabular}{@{}c@{}}Limited scope, \\ Inefficient load metric \end{tabular} & No & Only QoS \\
 & \begin{tabular}{@{}c@{}}SDN-based Architecture\\for Dense Networks \cite{reference020} \end{tabular} & Handover delay reduction & Limited study & Yes & Not stated \\
 \hline
\end{tabular}}
\end{center}
\end{table*}
\subsection{Balancing load in Access Networks}
\label{sec:2}
The works in~\cite{reference010,reference011,reference012,reference013,reference014,reference015,reference016,reference017} concentrate on the load management in the access network.~\cite{reference010} proposes LB strategies to optimally associate guaranteed bit rate (GBR) users to minimize their call drop and blocking rate (CDBR). However, this work only assumes the GBR users in the network. In~\cite{reference011}, the authors adopts a sequential strategy of handover, quality of service (QoS) aware scheduling, and admission control to achieve the designed objectives for GBR and Non-GBR end users in each uplink (UL)/downlink (DL) transmit time interval (TTI).~\cite{reference012} uses a similar approach as in~\cite{reference011}, however, it shows that the blocking rate of GBR/QoS users depends on the LB index as well as the network load. Thus, an aggregate objective function (AOF) for GBR user is defined, that weights the LB index and the network load to minimize the call blocking probability. Nonetheless, the selection of an optimal weighting factor is an open problem which is not addressed in~\cite{reference012}. Furthermore, note that based on the network environment an imbalance between the two parameters (e.g. via a higher weight value) significantly degrades user’s performance by immensely increasing one quantity over the other. In~\cite{reference013}, the authors suggests a new analytical model that can used to associate users based on different objectives e.g. achieved rate, throughput, and the delay. The model is theoretically shown to obtain optimal results for users with different QoS requirements. A dynamic handover and sub-band selection approach based on an online algorithm is suggested in~\cite{reference014}. The algorithm periodically accumulates and distributes the information about respective user gains to all network users. This work improves edge users performance and overall cell throughput while reduces the QoS violation probability when the network load is high. Authors in~\cite{reference027} design an approach to reduce the radio link failure due to high users handover rate. The new strategy increases the signaling between neighboring cells to adjust different handover parameters for each user dynamically. Note that these works not only ignores the backhaul network, but it has also high computational complexity due to the collection and distribution of user information for the final decision.

A new overloading factor is used for HetNets in~\cite{reference015} that manages the pico-cells coverage area based on the load of a macro cell. Thus the work shows that the pico-tier network performance can be optimized by carefully controlling the macro cell DL transmission. Similarly,~\cite{reference016} implements a control module at LTE evolved Node B (eNodeB), so that users can shift to a nearby location to enhance their spectral efficiency. The solution gives high network gain, but its user-driven nature limits its applicability in real network environments.~\cite{reference026} summarizes various types of handover schemes in HetNets focusing on users connectivity, load network-wide load balancing and energy efficiency. The work highlights that to achieve better performance, users can be switched based on extra information related to the states and the nature of target cells along with the received signal strength.~\cite{reference017} designs a framework for software-defined cellular network (SDCN), where the controller periodically computes delay of traffic flow at each BS by using the arrival and service rate information from each BS. This delay information then helps new users to select a suitable BS that provides better service rates, particularly for delay-sensitive traffic. Note that alongwith not utilizing the backhaul information the above works has one or more of the following differences: (1) they do not consider SDN, (2) their system performance is based on parameters that are impossible to determine apriori, (3) lack specific system level details and their experiments are not based on realistic network simulator.
\subsection{Load management in Backhaul Networks}
As shown in Table~\ref{tab:1}, for backhaul networks \cite{reference001,reference002} propose solutions that do not consider SDN. In~\cite{reference001} two load detection approaches are proposed in backhaul constrained LTE networks, first runs locally, while the second runs globally. The designed approach enables cells to dynamically change their network coverage for better LB. Results show that the local scheme can balance an individual BS scheduler load, while the global scheme enables different types of cells to adjust their coverage based on the available resources resulting in better load distribution across different cells. This work only assume elastic traffic and it does not evaluate the benefits achieved when both local and global LB approaches work together. A backhaul network congestion detection and control strategy is described in~\cite{reference002}, which periodically injects some packets into the network to avoid non-guaranteed bit rate (NGBR) packets to be randomly dropped upon backhaul congestion. The main drawback of this approach is the large overhead due to the probe packets sent between the gateway and the LTE eNodeB for congestion detection.

In~\cite{reference009,reference003}, SDN based backhaul network management is considered. A recent effort of SDN-based backhaul is made in~\cite{reference009}, where the authors proposed an OpenFlow based mechanism to prioritize traffic and control rate in LTE network. The essence of their work enables backhaul networks to guarantee the rate for GBR traffic during congestion due to the bottleneck link. Similarly, the work in~\cite{reference003} proposes an approach to tackle backhaul network congestion by utilizing OpenFlow. The main purpose of using a programmable backhaul is to allow network operators to share each others network infrastructure to achieve flexibility upon congestion.
\subsection{Load management in both Access and Backhaul Networks}
An enhanced mobility management approach in traditional LTE network is proposed in~\cite{reference018}, where the handover offset is tuned for the overloaded cells. A major setback of using it is that it presumes that the last mile link connecting the BS to the core network is the bottleneck. However, this is not always true, since the backhaul network usually has a ring topology~\cite{reference003} where the bottleneck can exist anywhere in the network. In~\cite{reference029} author outlines a framework that enables efficient traffic forwarding and QoS control within a wireless network by using SDN concepts.

Among the initial efforts for programmable cellular networks,~\cite{reference007} presents a SDCN design considering both the access and backhaul networks, simultaneously. Another recent work in~\cite{reference019} studies the impact of joint access-backhaul on mobility management solutions in 5G networks. Specifically for various QoS flows, different schemes can be used during cell selection based on the number of hops and the available links capacities.~\cite{reference020} presents a three tier SDN-based LTE network design for efficient mobility management. The results shows a significant delay reduction due to reduced signaling during handover event, however it lacks implementation on a system-level LTE simulator that enables to explore further benefits from the resulting integration.

In contrast to previous works the following are the key aspects of this work,
\begin{itemize}
  \item Handover objective for GBR and NGBR users are defined that results in better network performance using both access and backhaul networks.
  \item Backhaul transport network management system is developed for GBR and NGBR traffic in SDCN.
  \item We present system-level mobility LB procedures and describe specific information that need to be exchanged between local controller at BS and the central SDN controller for enhanced traffic management.
  \item Proposed framework is evaluated through extensive system-level simulations using ns-3.
\end{itemize}
\section{System Model}
\begin{table*}[t]
\scriptsize
\caption{All symbols and their definition}\centering
\label{tab:2}
\resizebox{\textwidth}{!}{
\begin{tabularx}{14cm}{|Sl|X|Sl|X|}
\hline
\textbf{Symbols}  & \textbf{Definition} & \textbf{Symbols} & \textbf{Definition} \\
\hline
$M$ & Set of all BSs & $\aleph$($i$) & Set of neighbouring cells of $i$ \\
\hline
$N$ & Set of total users in the network & $w^{i,j}_{net}$ ($t$) & Net available RBs at $i$ for $j$ \\
\hline
$N_{GBR}$ & Set of GBR users & $\xi$ ($t$) & Fairness index of load distribution \\
\hline
$N_{NGBR}$ & Set of NGBR users & $w^{i}_{AC} (t)$ & Available RBs at $i$ \\
\hline
$S_{b}$ & Set of SDN switches & $w^{i,j}_{BH} (t)$ & Available RBs at $i$ for $j$ w.r.t. backhaul \\
\hline
$N_o$ & Noise power spectral density & $SINR_{i,j,r}$ ($\tau$) & SINR of $j$ from $i$ over $r$ \\
\hline
$w^{i,j}_{GBR,used}$ & Used RBs of user $j$ of cell $i$ & $\overline{SINR}_{i,j}$ ($t$) & Avg. SINR of user $j$ from cell $i$ \\
\hline
$i$ & BS index & $j$ & User index \\
\hline
$r$ & Resource block index & $\tau$ & Subframe index \\
\hline
 $\ell$ & Length of LB period (in sec) & $h_{i,j,r}$ & Gain from $i$ to $j$ over $r$ \\
\hline
$D_{GBR, i}$ & Overall GBR users load & G(|$N_{NGBR}^{i}$|) & Multiuser diversity gain of NGBR users \\
\hline
$P^{tx}_{i,r}$ & Transmission power of $i$ & $\bar{\rho}_{GBR}$ ($t$) & Avg. network load of GBR users \\
\hline
$\eta_{i,j}$ ($t$) & Spectral efficiency of $j$ from $i$ & $I_{i,j}$ ($t$) & Association indicator of $j$ with $i$ \\
\hline
$d^{j}_{GBR}$ & Rate demand of user $j$ & $w^{i}_{GBR, used}$ ($t$) & Used RBs at cell $i$ \\
\hline
$C^{i,j}_{BH}$ ($t$) & Backhaul rate of $i$ for user $j$ & $\rho^{i,j}_{GBR}$ ($t$) & Load of user $j$ on $i^{th}$ cell \\
\hline
$N_{NGBR}^{i}$ & Set of NGBR users served by $i$ & $U_{j}$ ($t$) & Utility of NGBR user $j$ \\
\hline
$C^{i}_{BH}$ ($t$) & Backhaul rate of cell $i$ & $R_{i,j}$ ($t$) & Rate of $j$ from $i$ \\
\hline
$RSRQ_{i, j}$ ($t$) & RSRQ of $j$ from $i$ & $G_{i,j}$ & GBR user $j$ gain when served by $i$ \\
\hline
$G^{j}_{i \rightarrow c}$ & Gain after switching to $c$ & $\delta_{GBR}$ & Threshold for GBR users gain \\
\hline
$RSRQ_{thresh}$ & RSRQ threshold & $h_{\Delta}$ & Hysteresis value \\
\hline
\end{tabularx}}
\end{table*}
This section presents the problem statement and the analytical foundation of our system. All indexes and symbols used in our model are defined in Table~\ref{tab:2}. In a multi-cell LTE frequency division duplex (FDD) network consisting of a set $M$ of all macro cells BSs, set $N$ of total users comprising of $N_{GBR}$ set of GBR users and $N_{NGBR}$ set of NGBR users, and $S_{b}$ set of SDN switches in the backhaul network. $S_{b}$ forms the core network data plane specifically containing the switches responsible for handling the serving gateway data plane (SGW-D) and packet data network (PDN) gateway data plane (PGW-D) functionalities as shown in Fig.~\ref{jpg2}b. For each cell $i$, we use $\aleph$($i$) as the set of neighbouring cells of $i^{th}$ cell. When a new user arrives in the network, the relevant initial context setup requests and response messages are exchanged between the BS and the MME. This informs the MME about the related cell identifier, evolved radio access bearer identifier (ERAB-ID), QoS class identifier (QCI), and other information about the new arrived user. The SDN controller, being a part of control plane gets all these user information from the MME as well. Further, the MME communicates with the SGW to inform the core network about the new users bearer service. Thus, a new bearer session is created connecting the user to the IP-based network through the evolved packet core (EPC).
\subsection{Link Model}
To define the load contributed by the respective GBR or NGBR user in the presence of available backhaul and access networks resources, we first start from the SINR model. In a macro cell network environment, the instantaneous SINR at an end user $j$ from cell $i$ over resource block (RB) $r$ within a subframe $\tau$ is formulated as,
\begin{equation}\label{eq01}
  SINR_{i,j,r} (\tau) = \frac{P^{tx}_{i,r} (\tau) \cdot |h_{i,j,r}|^{2}}{ \sum_{\forall k \epsilon M, k \neq i} P^{tx}_{k,r} \cdot |h_{k,j,r}|^{2} + BW \cdot N_o}
\end{equation}
In above, $P^{tx}_{i,r}$ and $h_{i,j,r}$ defines the transmission power of the $i^{th}$ macro cell and the channel gain between cell $i$ and user $j$ over RB $r$, respectively. $\sum_{\forall k \epsilon M, k \neq i}$ $P^{tx}_{k,r}$ $|h_{k,j,r}|^{2}$ is the interference from all other cells other than the serving cell to user $j$ (as the frequency reuse is 1), $N_o$ is noise power spectral density, and BW is the RB bandwidth. If $\overline{SINR}_{i,j}$ ($t$) defines the average SINR of all RBs at time $t$ within sub-frame $\tau$ (i.e. $\tau$ $\in$ ($t$-$\ell$, $t$)). Then the achieved spectral efficiency of user $j$ from $i^{th}$ cell over all sub-frames $\tau$ will be $\eta_{i,j}$ ($t$), and is the log of the average SINR i.e. $\log_2$(1 + $\overline{SINR}_{i,j}$ ($t$)) from Shannon's capacity theorem~\cite{reference022}. In above, note that $\ell$ is the length (in seconds) of LB period.
\subsection{Problem Definition}
For robust mobility management, the inter-cell association decision is made by the respective BS. This often yields sub-optimal results  because of the distributed nature as the information available at each BS is limited. SDN enables network programmability by making the network architecture more simple and manageable. This work uses SDN concepts to achieve an efficient LB that not only consider both the access and backhaul networks but also utilize different mechanisms for GBR and NGBR traffic.
\subsubsection{GBR Users}
The load due to GBR users is the ratio of the number of resources they occupy to the total number of available resources in the network. We define the overall load due to GBR users on cell $i$ i.e. $D_{GBR, i}$ as $\sum_{\forall j \epsilon N_{GBR}}$ $I_{i,j}$ ($t$) $\cdot$ $d^{j}_{GBR}$. Here $I_{i,j}$ ($t$) is an association indicator, it is 1 if user $j$ is associated to the $i^{th}$ BS over $t$, otherwise it is 0. $d^{j}_{GBR} (t)$ is the DL rate requirement of $j^{th}$ GBR user at time $t$. From here on, we will use notations $AC$ and $BH$ to signify access and backhaul networks, respectively.

In a cell $i$, $w^{i}_{GBR, used} (t)$ defines the used time-frequency RBs at time $t$. The fraction of used RBs by GBR user $j$ served by cell $i$ is,
\begin{equation}\label{eq02}
  w^{i,j}_{GBR,used} (t) = \frac{I_{i,j} (t) \cdot d^{j}_{GBR} (t)}{\min{(BW \cdot \eta_{i,j} (t), C^{i,j}_{BH} (t))}}
\end{equation}
where $C^{i,j}_{BH} (t)$ is the backhaul supported link rate of cell $i$ for user $j$ at time $t$. Further $w^{i}_{GBR,used} (t)$ is the total RBs used by all GBR users of cell $i$ and is define as $\sum_{\forall j \in N_{GBR}} I_{i,j} (t) \cdot w^{i,j}_{GBR,used} (t)$.

It can be seen that the capacity of the access network is limited by the available RBs and the spectral efficiency of the $i^{th}$ cell while for backhaul network it is limited by the supported link rate for cell $i$. The load of the $i^{th}$ cell due to GBR user $j$ (i.e. $\rho^{i,j}_{GBR}$ ($t$)) is the ratio of the number of occupied resources to the number of available resources that could either be from access ($w^{i}_{AC} (t)$) or backhaul ($w^{i,j}_{BH} (t)$) networks and is given as,
\begin{equation}\label{eq03}
  \rho^{i,j}_{GBR} (t) =  \frac{w^{i,j}_{GBR,used} (t)}{\min{(w^{i}_{AC} (t), w^{i,j}_{BH} (t))}}
\end{equation}
where $w^{i}_{AC} (t)$ is the number of available RBs at cell $i$ and $w^{i,j}_{BH} (t)$ is the number of available RBs in access network for user $j$ as seen by the backhaul and is given as $\lceil\frac{C^{i}_{BH} (t)}{BW \cdot \eta_{i,j} (t)}\rceil$. Here $\lceil$a$\rceil$ is the minimum integer value larger than $a$. $C^{i}_{BH} (t)$ is the backhaul link available bandwidth of cell $i$ at time $t$. Note that if the backhaul supported rate from cell $i$ is large and spectral efficiency is low, this dictates a general condition of interference-limited networks, where the user resources in access network limits the user rate ($w^{i}_{AC} (t)$ $<$ $w^{i,j}_{BH} (t)$). Similarly for backhaul constrained networks,  $w^{i,j}_{BH} (t)$ represents the available access network resources over $t$ for user $j$ as seen from backhaul ($w^{i,j}_{BH} (t)$ $<$ $w^{i}_{AC} (t)$). For simplicity, we use $w^{i,j}_{net} (t)$ for $\min{(w^{i}_{AC} (t), w^{i,j}_{BH} (t))}$ as the net available RBs at cell $i$ for user $j$.

Using Eq.~\ref{eq03} the total GBR load for BS $i$ at time $t$ is $\sum_{\forall j \in N_{GBR}}$ $I_{i,j} (t)$ $\cdot$ $\rho^{i,j}_{GBR} (t)$. Consequently, the average load due to GBR users is,
\begin{equation}\label{eq04}
  \bar{\rho}_{GBR} (t) = \frac{\sum_{\forall i \epsilon M} \rho_{GBR, i} (t)}{|M|}
\end{equation}
In our network settings, we use Jain's fairness index~\cite{reference023} to define the fairness in user load distribution between macro cells as,
\begin{equation}\label{eq05}
  \xi (t) = \frac{(\sum_{\forall i \epsilon M} \rho_{GBR, i} (t))^{2}}{ |M| \cdot \sum_{\forall i \epsilon M} {(\rho_{GBR, i} (t))^{2}} }
\end{equation}
where $\xi (t)$ ranges from [$\frac{1}{|M|}$, 1], the larger it is, the more balanced would be the users between BSs.
\subsubsection{NGBR Users}
For NGBR users, the goal is to select the target cell that maximizes the resulting network utilization. This means how efficiently network resources can be used to enhance the achievable rate of all NGBR users. As utility of a NGBR user $j$ from cell $i$ is a monotonically increasing function of achievable data rate, which can be calculated in case of proportional fair scheduling similar to~\cite{reference011,reference012} as,
\begin{equation}\label{eq06b}
   R_{i, j} (t) = \eta_{i,j} (t) \cdot BW \cdot \lfloor\frac{\overline{w}^{i,j}_{net} (t)}{|N_{NGBR}^{i}|}\rfloor \cdot G(|N_{NGBR}^{i}|)
\end{equation}
where $\lfloor$a$\rfloor$ is the maximum integer value smaller than $a$ and $\overline{w}^{i,j}_{net} (t)$ is the net available RBs for NGBR user $j$ over $t$ and is given by $w^{i,j}_{net} (t)$ - $w^{i}_{GBR,used} (t)$. $|N_{NGBR}^{i}|$ is the number of NGBR users served by BS $i$. Also $G$($|N_{NGBR}^{i}|$) is the multi-user diversity gain of $|N_{NGBR}^{i}|$ users served by cell $i$ and is calculated as $\sum_{\forall j \in N_{NGBR}^{i}}$ $\frac{1}{j}$ from~\cite{reference012}. Further note that $\frac{w^{i,j}_{net} (t) - w^{i}_{GBR,used} (t)}{|N_{NGBR}^{i}|}$ $\cdot$ $G(|N_{NGBR}^{i}|)$ signifies the fraction of RBs used by NGBR user $j$ in case of proportional fair scheduling.
\subsubsection{Objective Function}
Maximizing fairness for the GBR users though equalizes the resource utilization among cells, but such an association is unaware of users channel condition. Thus, it degrades the overall network performance as user may be associated to a far away BS that results in lower spectral efficiency. The goal should be to consider both the distribution of resources consumption across different eNodeBs as well as the actual resources consumed by eNodeBs in the network. The later captures the users channel condition, since a GBR user (having fix rate requirement) with a better channel will require fewer time/frequency resources to satisfy its rate compared to the GBR user with a bad channel. Thus, when users handover to other eNodeBs, it not only changes the distribution of resource consumption across different eNodeBs but it also affects the resources consumed by different eNodeBs (/BSs) in the network. To summarize this mean that the load represented via Eq.~\ref{eq03} changes with the eNodeB $i$.

Similar to~\cite{reference016}, where different parameters for the GBR user’s utility are combined in a multiplicative way, we used the product of fairness index ($\xi (t)$) and proportion of average available resources (1 - $\bar{\rho}_{GBR} (t)$~\footnote{$\bar{\rho}_{GBR} (t)$ is the average network load that reflects the channel metric.}) as our objective. In the literature, different functions are used for optimizing the GBR user’s performance. For example, in~\cite{reference010,reference011} authors only used fairness index to allocate users in a multi-cell network. Also~\cite{reference016} uses the product of spectral efficiency and the available network resources over an eNodeB for handover decision. An obvious way to combine parameters is in~\cite{reference012}, where a weighted sum of fairness and average network load, based on a weight-age parameter is used. Note that, the work uses the case of asymmetric distribution of stationary users across cells in multi-cellular networks and assumes that optimal weight-age value is-known in advance. However, in a real multi-cell network environment with random user mobility, finding an optimal weight between fairness and load is not trivial as it depends upon several factors (e.g. channel conditions, user mobility, etc.). Further, we found that based on the network environment an imbalance between the two parameters (with a higher or lower weight value) significantly degrades either the user's throughput or the overall network fairness in resource usage by eNodeBs. Therefore, we need to update both quantities with respect to one another without significantly demeaning one or the other.

The essence of our objective is to perform a handover if the resulting increase in fairness in load distribution exceeds the increase in the average network load, or vice versa. If $\xi_{old} (t)$, $\xi_{new} (t)$ and $\bar{\rho}^{old}_{GBR} (t)$, $\bar{\rho}^{new}_{GBR} (t)$ define the fairness indexes and average network loads, respectively, before and after the handover, we can state the condition as,
\begin{equation}\label{eq07}
  \frac{\xi_{new} (t)}{\xi_{old} (t)} \geq \frac{1 - \bar{\rho}^{old}_{GBR} (t)}{1 - \bar{\rho}^{new}_{GBR} (t)}
\end{equation}
Next, if $R_{i,j} (t)$ is the rate of user $j$ in access network and $w^{i}_{BH} (t)$ is the backhaul resources available in cell $i$, then the overall problem for GBR users is,
\begin{subequations}
\begin{alignat}{3}
\! & \max_{\mathbf{I_{i,j} (t)}} \xi (t) \cdot (1 - \bar{\rho}_{GBR} (t)),~such~that \label{eq100} \\
\! & \sum_{\forall i \in M}I_{i,j} (t) \cdot min(R_{i,j} (t), C^{i,j}_{BH} (t)) \geq d^{j}_{GBR} (t), \forall j \in N_{GBR}\label{eq101} \\
\! & \sum_{\forall j \in N_{GBR}}I_{i,j} (t) \cdot w^{i,j}_{GBR, used} (t) \leq min(w^{i}_{AC} (t), w^{i}_{BH} (t)), \forall i \in M\label{eq102} \\
\! & \sum_{i \epsilon M} I_{i,j} (t) = 1, \forall j \in N_{GBR}.\label{eq103}
\end{alignat}
\end{subequations}
Similarly in~\cite{reference001}, we know that the traffic load due to NGBR users is inversely proportional to the achieved rate limit. Thus, for NGBR users in the network $N_{NGBR}$ the optimization problem is stated as,
\begin{subequations}
\begin{alignat}{3}
\! & \max_{\mathbf{I_{i,j} (t)}} \sum_{\forall i \in M} \sum_{\forall j \in N_{NGBR}} U_{j} (I_{i,j} (t) \cdot R_{i,j} (t)),~such~that\label{eq111} \\
\! & \sum_{\forall i \epsilon M} I_{i,j} (t) = 1, \forall j \in N_{NGBR}\label{eq114} 
\end{alignat}
\end{subequations}
Eq.~\ref{eq100} and Eq.~\ref{eq111} defines the handover goals for the GBR and NGBR users, respectively. This means to select the user association matrix $I_{i,j}$ $\forall$ $i$ $\in$ $M$ and $j$ $\in$ $N$ that maximizes the above objective functions. For GBR users, the criterion is to select the target cell that evenly distributes users across the network while minimizing the average network load. The Eq.~\ref{eq101} and~\ref{eq102} signifies the rate constraint and available resources constraint for GBR users and cells ($i \epsilon M$), respectively. Eq.~\ref{eq103} and~\ref{eq114} define the association constraints for GBR and NGBR users respectively.

Due to the inherent complexity of the above problem, as well as the non-linearity introduced by the limited backhaul, solving it globally is infeasible as it would be computationally expensive, and will cause unnecessary delay and overhead. We use a distributed approach (further explained in Section 4) that divides the global problem into several sub problems, each one solved locally at an eNodeB. This simplifies the optimization and reduces the range of constraints set, representing the domain of our problem. Whereas, ignoring the resulting complexity and latency, the solution of global optimization problem can be used as an upper bound. 

To solve the problem in a distributed fashion, we adopt procedures of user scheduling, QoS-aware handover, admission, and rate control decisions to achieve the desired goals defined above. User scheduling and admission control decisions are performed in a distributed fashion for each wireless and backhaul network segment by the BS and SDN controller respectively. Using our SDN-LTE network architecture, we consider that a controller is implementing control procedures for managing a backhaul network of a set of macro cells. Each BS keeps track of the backhaul bottleneck available capacity notified by the controller. Every time a GBR flow is updated (added/modified/removed) from the specific bottleneck link in the backhaul, the controller reports the backhaul network capacity to the corresponding BS and its neighboring BSs, as it has complete information of the overall network. Using Eq.~\ref{eq03}, the BSs calculate the load contributed by each served GBR user and the respective overall loads within the LB period $\ell$, using the available access network RBs and backhaul capacity information. In the resulting SDN-LTE network, each BS has up-to-date information about the available backhaul bandwidth of all neighbouring BSs, which will be used to make efficient handover decision. After the handover decision is confirmed by the end user (shown via procedure 6 in Fig.~\ref{jpg3}), the SDN controller gets notified from the target BS via the MME. Later on, the controller will configure the network by adding/modifying corresponding flow table entries in the SDN switches.
\subsection{Assumptions}
A flat fading channel model is used. Similar to a typical LTE network, all neighboring BSs (/eNodeBs) are connected via X2 interfaces, enabling X2 control signaling between BSs. Hence the relevant resource status request, resource status response, and resource status update message information are exchanged over this interface. Finally, in our setup the controller is close enough to the switches enabling instant flow tables upgrade and rule installation, when required. Note that in our setup the available backhaul capacity is equally divided among the UL/DL traffic flows.
\section{Load Balancing With Both Wireless and Backhaul}
This section covers the LB mechanisms consisting of user scheduling, handover, admission and rate control procedures for GBR and NGBR traffic.
\subsection{Users Scheduling}
DL users traffic scheduling is done using the channel and QoS aware (CQA) scheduler~\cite{reference024}. At a corresponding eNodeB, this scheduler assign RBs to users flows based on the types of traffic (e.g. GBR/NGBR), the head-of-line (HOL) delay, and the channel quality over different sub-bands. Note that the CQA scheduler consists of the joint time-domain (TD) and frequency-domain (FD) scheduling done sequentially, for the users traffic. TD scheduling groups users flows for each TTI based on their HOL delay. In this case, the group with the highest HOL delay will be served first. The FD scheduler starts with assigning the available RBs in each TTI to the groups of flows forwarded to it by the TD scheduler. The assignment of RBs is done using the combined metric consisting of HOL delay, the desired bit rate (for GBR flows), and a channel aware metric. In our work, we considered that the channel aware metric selected based on the proportional fair metric that assign weights to users flows based on the ratio of their expected achievable throughput over the given RB and the past average throughput. 
\subsection{Load Estimation for GBR Users}
First, we describe the load estimation procedure that enables the serving BS to preemptively estimate the load of new users offered to the target BS. Later on, utilizing the available information gathered at each serving BS from the target BS via the X2 interface, and from the controller, we describe the handover decision criteria for GBR and NGBR users respectively.

The serving BS utilizes the handover user received power and the overall load information to estimate the user load over target BS after handover. From Eq.~\ref{eq01}-~\ref{eq06b}, we note that the achievable rate is directly proportional to the received power at the end user, and inversely proportional to the resulting load. Therefore, with current load information at the serving BS together with the reference signal quality (RSRQ) of the serving and target cells, the serving BS $i$ finds the resulting load of end user $j$ over the target macro cell. For a target BS $k$ $\in$ $M\setminus{i}$, the used RBs for user $j$ (i.e. $w^{k,j}_{GBR,used} (t)$) can be estimated as $\frac{w^{i,j}_{GBR,used} (t) \cdot RSRQ_{i, j} (t)}{RSRQ_{k, j} (t)}$. Using the fraction of used RBs for the new user and the net available RBs, we can find the load of user $j$ at BS $k$ as $\frac{w^{k,j}_{GBR,used} (t)}{w^{k,j}_{net} (t)}$.
\subsection{Handover Decision}
We now describe different handover decision criteria for GBR and NGBR users, considering their specific goals (Eq.~\ref{eq100} and Eq.~\ref{eq111}). As shown in the handover signaling flow in Fig.~\ref{jpg3}, the BS uses the received power and backhaul network information to make a final handover decision.
\subsubsection{For GBR Users}
The handover gain for GBR users can be found comparing the objective given in Eq.~\ref{eq100} before and after handover (as shown in Eq.~\ref{eq07}). This means that if before/after the handover the network objectives are defined by $G_{i,j}$/$G_{c,j}$ (where $c$ $\in$ $\aleph(i)$), respectively. Then the resulting handover gain $G^{j}_{i \rightarrow c}$ i.e. $\frac{G_{c,j}}{G_{i,j}}$ achieved by switching user to target cell $c$ $\in$ $\aleph(i)$ will be,
\begin{equation}\label{eq121}
  G^{j}_{i \rightarrow c} = \frac{y \cdot (x+\triangle)^{2} \cdot (|M| - x - \triangle)}{z \cdot x^{2} \cdot (|M| - x)}
\end{equation}
In above, $\triangle$ $=$ $\rho^{j}_{GBR, c} (t)$ - $\rho^{j}_{GBR, i} (t)$, $x$ $=$ $\sum_{\forall m \in M}$ $\rho_{GBR, m} (t)$, $y$ $=$ $\sum_{\forall m \in M}$ $\rho^{2}_{GBR, m} (t)$, and $z$ is same as $ \sum_{\forall m \in M\setminus{i,c} }$ $\rho^{2}_{GBR, m} (t)$ $+$ ${(\rho_{GBR, i} (t) - \rho^{j}_{GBR, i} (t))}^{2} + {(\rho_{GBR, c} (t) + \rho^{j}_{GBR, c} (t))}^{2}$. From the above equation, the serving cell $i$ select the cell $k$ among $c$ $\in$ $\aleph(i)$ cells for which the handover gain $G^{j}_{i \rightarrow k}$ is maximum and  $G^{j}_{i \rightarrow k}$ $>$ 1+$\delta_{GBR}$ is true. Note that the threshold $\delta_{GBR}$ depends on the network environment to minimize the ping-pong effect. 

After the handover decision is made the target BS will be notified as in Fig.~\ref{jpg3}, which then perform admission control for the newly arriving user flow.
\subsubsection{For NGBR Users}
As mentioned earlier, a NGBR users need to be associated to the BS from whom it gets the maximum rate. For the total number of $|N_{NGBR}|$ NGBR users in the network, the handover condition when a NGBR user $j$ moves from cell $i$ to $k$ is defined $\forall$$j$ $\in$ $N_{NGBR}$ as to choose cell $k$ for which the user gets best rate i.e. $R_{k,j} (t)$ $\geq$ $R_{i,j} (t)$, where $k$ $\neq$ $i$. 

In the presence of NGBR traffic, the antenna sub carrier utilization factor for cell in full load conditions will be $\frac{5}{6}$. As SINR is an increasing function of received signal power information over the whole OFDM symbol (e.g. RSRQ). Thus the spectral efficiency $\eta_{k,j} (t)$ can be estimated using $RSRQ$. Then, with the used resources information the achievable rate of $j^{th}$ user from cell $k$ can be calculated. Hence using the available backhaul rate information the handover condition for all NGBR users $\forall$$j$ $\in$ $N_{NGBR}$ is defined from Eq.~\ref{eq06b} as,
\begin{align}\label{eq13}
&RSRQ_{k,j} \cdot \frac{\overline{w}^{k}_{net} (t)}{|N_{NGBR}^{k}| + 1} \cdot G(|N_{NGBR}^{k}|+1) \geq \nonumber\\
&RSRQ_{i,j} \cdot \frac{\overline{w}^{i}_{net} (t)}{|N_{NGBR}^{i}| - 1} \cdot G(|N_{NGBR}^{i}|-1)
\end{align}
From above, $\overline{w}^{k}_{net} (t)$ $=$ $w^{k}_{net} (t) - w^{k}_{GBR,used} (t)$ and $\overline{w}^{i}_{net} (t)$ $=$ $w^{i}_{net} - w^{i}_{GBR,used}$ are the residual resources of cell $k$ and $i$, respectively, left after the resources are allocated to the GBR users. Similarly, $N_{NGBR}^{k}(t) + 1$ and $N_{NGBR}^{i} (t) - 1$ are the number of NGBR users served by cell $k$ and $i$ over $t$ after the handover, respectively.
\subsection{User admission and rate control}
The admission control condition is checked by the BS for the respective GBR user, as shown in Fig.~\ref{jpg3}. This implies that the user $j$ is admitted by BS $k$ if it has enough resources to satisfy the user demand,
\begin{equation}\label{eq14}
  min(w^{k}_{AC} (t), w^{k,j}_{BH} (t)) - w^{k}_{GBR,used} (t) \geq w^{k,j}_{GBR,used} (t)
\end{equation}
In above, $w^{k,j}_{GBR,used} (t)$ is the RBs required to satisfy the rate demand of GBR user $j$ as defined by Eq.~\ref{eq02}, $w^{k}_{GBR,used} (t)$ is the total RBs used at BS $k$ by all of its existing GBR users, and min($w^{k}_{AC} (t)$, $w^{k,j}_{BH} (t)$) is the available RBs at BS $k$.

In the transport backhaul network, the available link bandwidth is divided between GBR and NGBR flows. For each admitted GBR user, our controller apply respective dedicated OpenFlow meter to control the associated traffic according to the rate requirement of the respective user as exchanged during initial bearer setup procedure. A GBR flow does not achieve the required rate if upon adding it, the aggregate capacity of GBR flows exceeds the available percentage of link capacity for GBR users. All NGBR flows in backhaul shares the fixed percentage of the available bottleneck link capacity using a single aggregate OpenFlow meter. Later we use a strategy where in case of congestion, a single aggregate meter is gets divided into multiple QoS based NGBR meters as explained in Section~\ref{section5c}.
\subsection{Traditional Handover Algorithms}
\subsubsection{Max-RSRQ based User Association}
In Max-RSRQ approach a user is associated to cell from whom it gets the best channel conditions. We use A2A4 RSRQ measurement algorithm~\cite{reference025}, which depends on A2 and A4 measurement reports sent by the end user to its serving cell. In this case, a specific $RSRQ_{thresh}$ is used by all users to match their received DL signal RSRQ value against it. An A2 report signifies that the serving cell RSRQ is below this threshold, while an A4 report says that neighbor RSRQ values become better than the RSRQ threshold. If $h_{\Delta}$ is the hysteresis value and $RSRQ_{i,j}$ is the signal strength received at user $j$ from BS $i$, then the overall handover condition is defined by,
%
\begin{equation}\label{eq15}
\resizebox{1.0\hsize}{!}{$(RSRQ_{k,j} > (RSRQ_{i,j} + h_{\Delta})) \wedge (RSRQ_{k,j} > RSRQ_{thresh} \vee RSRQ_{i,j} < RSRQ_{thresh})$}
\end{equation}
\subsubsection{QoS-aware Handover Algorithm}
The above handover scheme is unaware of the resulting achievable throughput and may result in overloading some BSs compared to others, especially in asymmetric users distribution scenarios. Hence new cell association algorithms are designed that not only considers the received signal power but also uses the load information. Different previous works discusses load-aware handover algorithms, however their work is only limited on achieving fairness in load distribution within access network for GBR users. Therefore as a comparative LB scheme we use the QoS aware handover version~\cite{reference011}, where the fairness in load distribution ($\xi (t)$) is considered for GBR users as defined by Eq.~\ref{eq05}.
\subsection{Proposed Optimized Handover Algorithm}
\begin{figure}[t!]
\begin{center}
\resizebox{3.5 in}{!}{\includegraphics{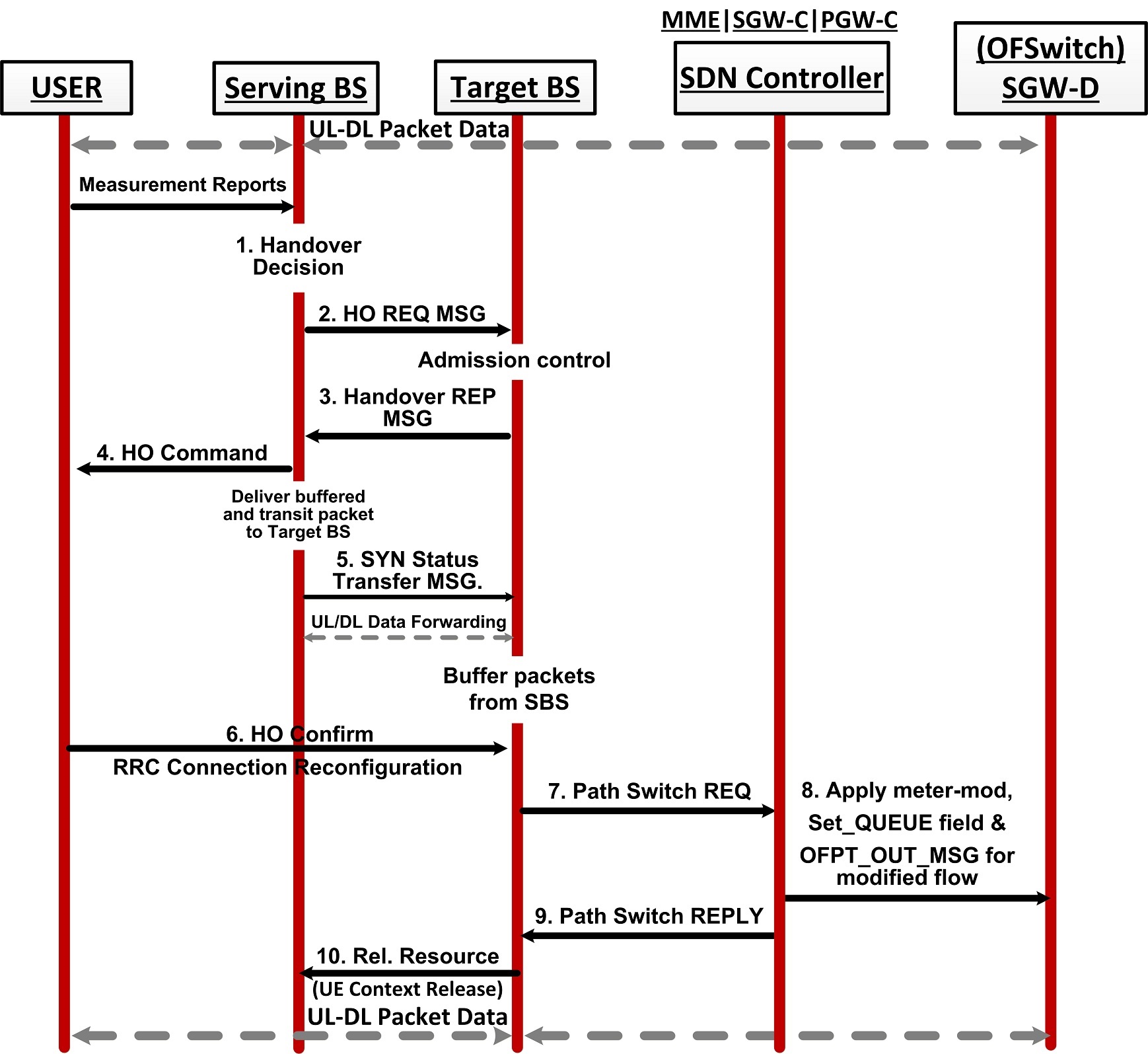}}
\caption{\em Handover procedure in proposed SDN based LTE network ~\cite{reference004}}
\label{jpg3}
\end{center}
\end{figure}
Eq.~\ref{eq100}-\ref{eq103} and Eq.~\ref{eq111}-\ref{eq114} presents global optimization problems and their respective constraints for GBR and NGBR users, respectively. Solving these problems globally is inefficient due to resulting complexity and latency in acquiring and processing information from eNodeBs at a central unit. For instance, if an exhaustive search is use as a global solution, it will exponentially increase the complexity, as in that case one need to evaluate the objective for all of the $|M|^{|N|}$ outcomes. Where |M| and |N| represents the number of eNodeBs and the number of users, respectively. 

Since LTE has a flat network architecture, we use a distributed approach by dividing the global problem into several sub problems, each one solved locally at an eNodeB. Each local optimization problem has the same objective as of the original global problem, with reduced range of constraints set (e.g. number of eNodeBs and users), which represents the domain of the problem. As an example, for each serving BS the candidate set of users corresponds to the current number of users served while the candidate BSs are the neighboring BSs (from measurement reports) to whom the served user can switch. The locally optimal solution compares the objective (in Eq.~\ref{eq100} and Eq.~\ref{eq111}) before and after the handover. The results via Eq.~\ref{eq121} and Eq.~\ref{eq13} (described in Section 4.3.1 and 4.3.2) represents the handover gain from the locally optimal solution when a GBR/NGBR user $j$ is switched from an eNodeB $i$ to the neighboring eNodeB $c$. Thus, each serving BS uses its own load information, neighboring eNodeBs load, and the backhaul network information to make a decision by solving the problem locally.

Algorithm~\ref{table:algorithm1} shows our proposed algorithm that solves the local optimization problem at each eNodeB. From line~\ref{line11} in Algorithm 1, the handover procedure triggers when the serving BS receive a handover request from user due to low signal quality. At time $t$, for the current load balancing (LB) period ($l$), the BS uses the overall collected user’s measurement reports, backhaul network resources, and the users load information to calculate handover gain for the GBR or NGBR users, respectively. Based on the handover gain, the serving BS makes handover decision (as in Fig.~\ref{jpg3}) to select and connect to the target BS. For example, in case of GBR users the target BS $k$ and user $j$ are selected that maximizes the gain condition as described in Section 4.3.1. Once a target BS is chosen, the controller gets notified, which manages active bearer statistics and perform the procedures summarized in algorithm~\ref{table:algorithm1}. At first, the controller installs flow rules to establish the particular route based on the routing policy. Then it apply priority based scheduling and rate control for GBR and NGBR users to guarantee the available bandwidth for the GBR traffic in the backhaul network. The procedures for our software-defined backhaul management are explained in the next section.
\begin{algorithm}
\caption{Proposed Handover Algorithm for SDN-based LTE Networks}\label{table:algorithm1}
\begin{algorithmic}[1]
\State{\textbf{Symbols and Functions}}
\State Indexes $i$, $k$, and $c$ for BSs, index $j$ for end user \label{line00}
\State $\aleph(i)$ $\gets$ \text{Set of neighbouring BSs of BS $i$}  \label{line01}
\State $N_{GBR}$ $\gets$ \text{Set of GBR users served by the network} \label{line02}
\State $N_{NGBR}$ $\gets$ \text{Set of all NGBR users served by the network} \label{line03}
\State $G_{i,j}$ $\gets$ \text{$j^{th}$ GBR users gain when connected to BS $i$} \label{line04}
\State $C^{k,j}_{BH}$ ($t$) $\gets$ \text{Available backhaul capacity for $k$ $\in$ $\aleph(i)$ BSs for $j^{th}$ users flow} \label{line05}
\State $RSRQ_{thresh}$ $\gets$ \text{minimum RSRQ threshold value}  \label{line07}
\State $\rho_{GBR,k}$ ($t$) $\gets$ \text{GBR users load on BS $k$} \label{line08}
\State \textmd{Estimate-Load(i,c,j)} $\gets$ \text{Function estimates the load $\rho^{j}_{GBR, c}$}\label{line09}
\State \textmd{Connect(j, i)} $\gets$ \text{Function to connect user $j$ to target BS $i$}  \label{line10}
\end{algorithmic}
\end{algorithm}
\begin{algorithm}
\begin{algorithmic}[2]
\Procedure{ At Base Station (BS) $i$}{}
\If{($\textit{RSRQ}_{j, i} < \textit{RSRQ}_{thresh}$)} \label{line11}
\myState{RECV(\textbf{HO-REQUEST}) from user} \label{line12}
\myState{\textbf{Decision Engine}} \label{line13}
\myState{\If{($j$ $\epsilon$ $N_{GBR}$)}} \label{line14}
\myState{Get back-haul resource information}
\myState{Calculate load $\rho_{GBR,k} (t)$, $\forall$ $k$ $\in$ $\aleph(i)$} \label{line15}
\For{c = 1; c $\leq$ $|\aleph(i)|$; ++c} \label{line16}
\myState{\textmd{Estimate-Load(i,c,j)}} \label{line17}
\myState{Calculate $G_{c,j}$ and $G_{i,j}$ using Eq.~\ref{eq100}} \label{line18}
\myState{Find $G^{j}_{i \rightarrow c}$ via Eq.~\ref{eq121}} \label{line19}
\If{(($G^{j}_{i \rightarrow c}$ $>$ $1+\delta_{GBR}$) $\wedge$ ($G^{j}_{i \rightarrow c}$ $>$ $G^{j}_{i \rightarrow k}$))} \label{line20}
\myState{Choose $k$ $=$ $c$}  \label{line21}
\EndIf  \label{line22}
\EndFor   \label{line23}
\myState Connect(j, k) as in pro.~2 of Fig.~\ref{jpg3} \label{line24}
\ElsIf{($j$ $\epsilon$ $N_{NGBR}$)} \label{line25}
\If{( HO criteria for NGBR satisfied)} \label{line26}
\myState{Select target BS $k$ maximizing the user rate using Eq.~\ref{eq13}} \label{line27}
\myState{Connect(j, k)} \label{line28}
\EndIf \label{line29}
\EndIf \label{line30}
\myState{\textbf{HO-REPLY} from target BS in pro.~3 Fig.~\ref{jpg3}} \label{line31}
\myState{Forward \textbf{HO-COMMAND} as in pro.~4 Fig.~\ref{jpg3}} \label{line32}
\myState{\Else } \label{line33}
\myState{No Handover } \label{line34}
\EndIf \label{line35}
\EndProcedure \label{line36}
\end{algorithmic}
\end{algorithm}
\begin{algorithm}
\begin{algorithmic}[3]
\Procedure{ At SDN Controller}{}
\myState{Keep all active bearers statistics up-to-date} \label{line37}
\myState{RECV(\textbf{Path-Switch-REQ}) as in pro. 7 of Fig.~\ref{jpg3}} \label{line38}
\myState{Select path (via shortest-path-first routing) to target BS $k$} \label{line39}
\If{($j$ $\epsilon$ $N_{GBR}$)} \label{line40}
\myState{Update link stats of available bottleneck capacity for user $j$, $C^{i,j}_{BH} (t)$ and $C^{k,j}_{BH} (t)$} \label{line41}
\EndIf \label{line42}
\myState{Apply priority based scheduling and rate control} \label{line43}
\myState Install flow rules with rate as in pro.~8 of Fig.~\ref{jpg3} \label{line44}
\EndProcedure
\end{algorithmic}
\end{algorithm}
\begin{figure}[t]
\begin{center}
\resizebox{4.0 in}{2.0 in}{\includegraphics{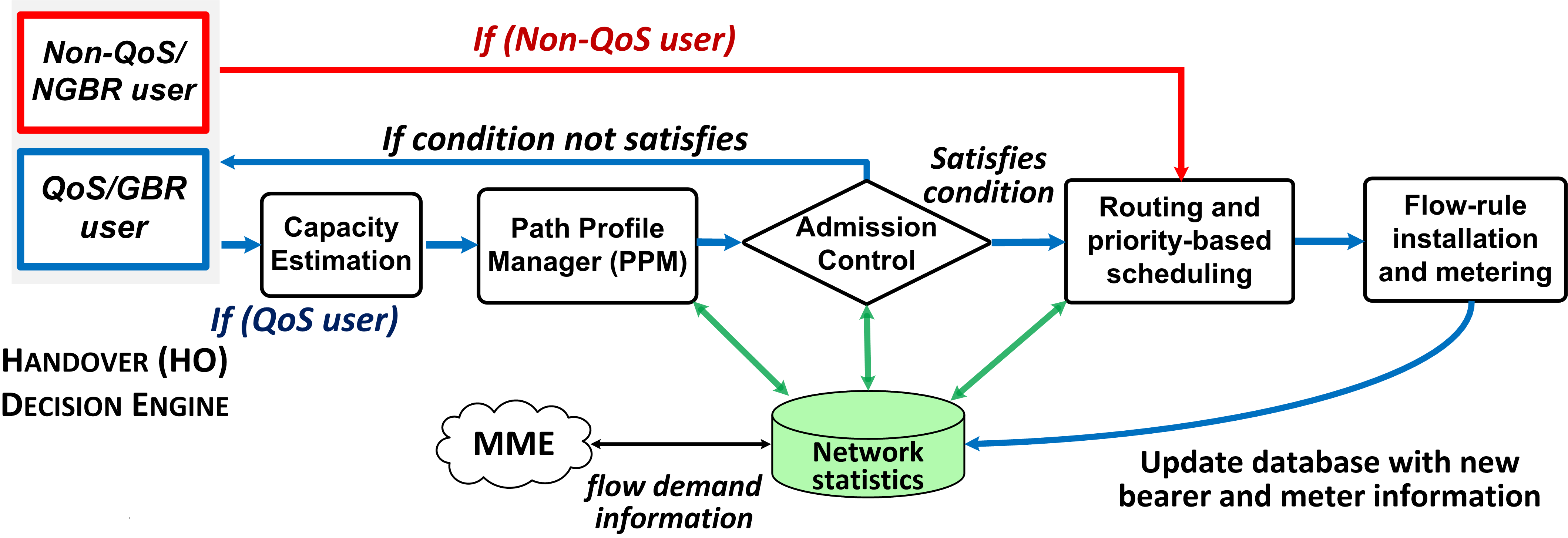}}
\caption{\em Control plane functionalities for backhaul network management. }
\label{jpg4}
\end{center}
\end{figure}
\section{Traffic Management in Backhaul Network}
The stepwise procedures performed for the GBR/NGBR flows in the backhaul network are shown in Fig.~\ref{jpg4}. Once a handover decision is made as in Fig.~\ref{jpg3}, the routing path is calculated based on the given policy. In the considered backhaul network of SDN switches and their corresponding interconnecting links, each link supports some limited bandwidth. For each $i^{th}$ macro cell BS, $L_{i}$ is the set of paths to the $i^{th}$ BS with unique backhaul links, similarly $p_{i}$ represents a path $p$ to BS $i$ in the backhaul network. Therefore the total backhaul bandwidth of BS $i$ can be defined as, $C^{i}_{BH} (t) = \sum_{\forall p \epsilon L_{i}} C_{min}^{i,p}$. Here $C_{min}^{i,p}$ is the bottleneck link capacity of path $p$ to $i^{th}$ BS (e.g. $p_{i}$) in the backhaul network. 
\subsection{Capacity Estimation}
When a new user flow (with the backhaul path $p_i$) is added to BS $i$, the residual bandwidth $C^{k}_{residual} (t)$ of all BSs $k$ $\in$ $S_p$ sharing the bottleneck path $p_i$ get reduced to $C^{k}_{residual} (t)$ = $C^{k}_{BH} (t)$ - $\sum_{\forall p \epsilon L_{i}} \sum_{\forall k \epsilon S_{p}} C^{k, p}_{used} (t)$. Where $C^{k, p}_{used}$ defines the used bandwidth of all other flows of BS $k$ over the path $p$.
\begin{table*}[t]
\caption{MATCH-ACTION Fields for SGW-D}\centering
\label{tab:3}
\resizebox{\textwidth}{!}{%
\begin{tabular}{|c|c|}
  \hline
  Match Fields & Action Fields \\
  \hline
  OXM-OF-ETH-TYPE=0x800, & OFPIT-METER:METER-ID \\
  \hline
  OXM-OF-IN-PORT=input port, & apply:OFPAT-SET-FIELD=OXM-OF-TUNNEL-ID:S5/S1 TEID NEW, \\
  \hline
  OXM-OF-TUNNEL-ID=(S1/S5 TEID OLD), & OFPAT-SET-FIELD=OXM-OF-IPV4-SRC:ip-src-new, \\
  \hline
  & OFPAT-SET-FIELD=OXM-OF-IPV4-DST:ip-dst-new, \\
  \hline
  & OFPAT-SET-FIELD=OXM-OF-ETH-SRC:eth-src-new,\\
  \hline
  & OFPAT-SET-FIELD=OXM-OF-ETH-DST:eth-dst-new,\\
  \hline
  & OFPAT-SET-FIELD=OXM-OF-OUT-PORT, \\
  \hline
  & OFPAT-SET-QUEUE=QUEUE-ID \\
  \hline
\end{tabular}}
\end{table*}
\subsection{Path profile manager and admission control}
For the users (/bearer) flow management in the backhaul, the network statistics database keeps the path of each flow in the backhaul network and parameters like total backhaul network bandwidth, residual backhaul bandwidth, used bandwidth, and the meter rate. Path profile manager shown in Fig.~\ref{jpg4}, selects the path according to the specified policy. Once a path gets sorted, all links are evaluated again in the updated network graph and those not satisfying the rate requirement (e.g. $d^{j}_{GBR}$) of GBR users are removed. Hence a GBR flow is admitted to the new cell $k$ if $d^{j}_{GBR}$ $\leq$ $C^{k}_{residual}$. Also in this work we adopted shortest-path-first (SPF) routing policy that finds the shortest available path from PGW to the specific BS in the DL. 
\subsection{Backhaul QoS control and monitoring module}
\label{section5c}
After a bearer satisfying the admission control is accepted and admitted, this module apply the corresponding meter and scheduling for the new/reconfigured flow in the backhaul network.

Next, we proposed congestion based traffic management of different QoS classes of NGBR flows. To achieve this, congestion-aware NGBR meters are implemented that divides the available capacity among NGBR flow classes with respect to their QoS factors. In general, when network congestion occur critical NGBR flows suffer from unfair packet losses in the backhaul as highlighted in~\cite{reference002}. In our approach, when the NGBR flows grows beyond some threshold in the backhaul-limited LTE network, the SDN controller detects it as a decrease in per NGBR flow available bandwidth and activates multiple QoS aware NGBR aggregate meters by splitting the single NGBR aggregate meter based on the QoS factor. For example, consider two types of NGBR  traffic flows (type A and B) e.g. HTTP and FTP. Even though type A has higher priority than type B, however the backhaul network treats them equally and thus randomly drops HTTP packets when congestion occurs. We apply different aggregate NGBR meters during network congestion, by giving a weight-age to each NGBR flow type as,
\begin{equation}\label{eq16}
  MR_{NGBR, l}^f =  \frac{Q_f}{\sum_{\forall j \in S^{NGBR}_l} Q_j} \cdot NGBR^{cap}_{l}
\end{equation}
where, $MR^f_{NGBR, l}$ defines the meter rate set for the respective NGBR flow type $f$ over link $l$ and $S^{NGBR}_l$ is the set of all types of NGBR flows sharing the network link $l$, and $Q_j$ is the QoS weight factor of the corresponding $j^{th}$ NGBR users flow. Hence by using the metering feature of OpenFlow 1.3 protocol, we can specify the meter rate $MR^f_{NGBR, l}$ for the NGBR flow type $f$ based on its weight-age factor $Q_f$.
\subsection{Flow rule installation}
Here, we install respective flow entries as flow-mod messages as shown in Table.~\ref{tab:3} to reconfigure the backhaul switches based on the routing path.
\subsection{Update Module}
After the flow rules are installed and the respective meter entries are set, in here the link parameters and users flow information are updated within the network statistics database. After updating the network statistics, the SDN controller notify the new backhaul bandwidth ($C^{i}_{BH} (t)$ and $C^{i}_{residual} (t)$) to the respective BS $i$ and all of its neighboring BSs $\aleph(i)$ as shown in Fig.~\ref{jpg2}. During the handover decision phase, the serving BS select a new target BS using this backhaul bandwidth information, such that the backhaul bandwidth satisfies the rate requirement of new GBR user, as described earlier.

\section{Experimental Setup}
\begin{figure}[t!]
\begin{center}
\resizebox{4.0 in}{3.4 in}{\includegraphics{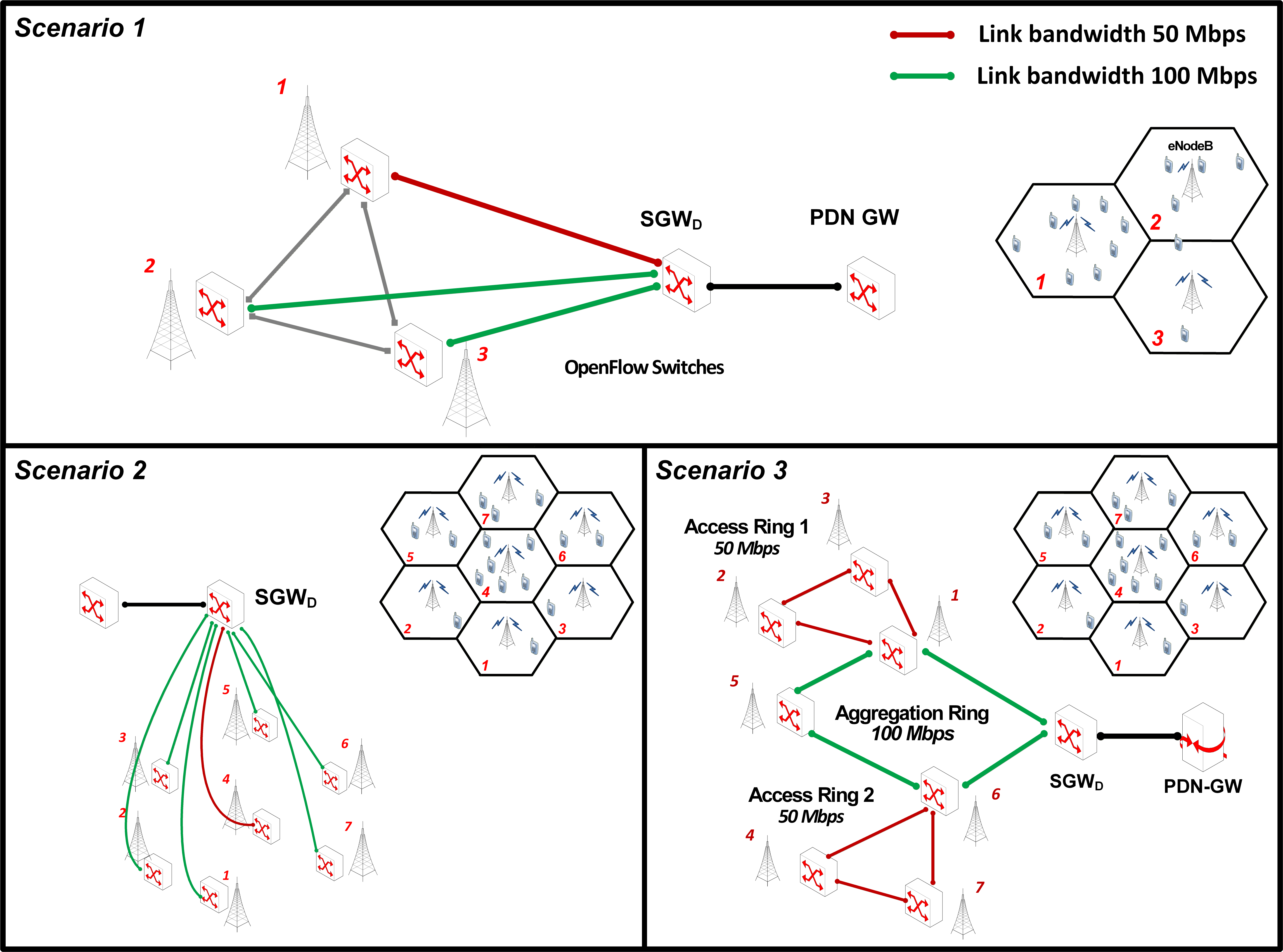}}
\caption{\em Different Scenarios for SDN-enabled LTE Networks. }
\label{jpg5}
\end{center}
\end{figure}
For our experiments, we consider hexagonal cell sites with 3 sectors per LTE eNodeB, and an inter-site distance of 500 meters. For the backhaul network consisting of SDN switches performing the SGW-D data-plane and backhaul network functionality, we modified the LTE module in ns-3. 
The modified SDN-enabled LTE module is basically an integration of the OpenFlow and LTE module in ns-3 (details in~\cite{reference004}). In this implementation, the MME works as a component of our customized SDN controller, and hence the controller is aware of all user sessions and their respective (QoS based) bearer mapping. The controller also emulates the functionality of the SGW-C (serving gateway control plane) by maintaining the S1-S5 bearer mapping at the OpenFlow switch, using the set-field command with the corresponding match-action fields (as shown in Table.~\ref{tab:3}).
\subsection{User Distribution}
Initially, we evaluated the system for an environment where all users are uniformly distributed over the coverage area. Then, we considered an asymmetric user distribution, where users are more densely concentrated within one of the BS coverage areas, compared to the rest. 
\subsection{User Mobility}
We consider an environment with majority (70\%) vehicular users~\cite{reference030} moving with an average speed of 30 km per hour. The rest 20\% users are walking at pedestrian speeds (0.9 m/s), while the remaining 10\% of users are stationary during the simulation. Users movement follows random walk mobility pattern implemented using the ns-3 class~\emph{RandomWalk2dMobilityModel} where the attribute pause time is set as 0.1 second.
For our experiments, we assume a Rayleigh distribution based fading model. The results are based on simulations with a duration of 30 seconds, while using realistic data channel error model available in ns-3 LTE module.
\subsection{NGBR/GBR Flows}
\subsubsection{Backhaul Capacity Distribution}
In our experiments, the capacity of each backhaul link is divided equally between uplink and downlink traffic. 
60\% of the capacity of each backhaul link is reserved for GBR flows, and the remaining 40\% are available for NGBR flows. For each GBR flow, a dedicated meter is used that limits the rate to a maximum of 250 kbps.
All NGBR flows are sharing their allocated backhaul link capacity using an aggregate meter. When the number of GBR flows increases beyond the available capacity, and no path is found from the PGW to the serving BS, the corresponding flows are simply dropped.
\subsubsection{NGBR/GBR Users Percentage}
Since we are mainly interested in evaluating the network performance for QoS users, we are using a ratio of the number of GBR to NGBR users of 9 to 1, unless otherwise specified.
\subsection{Experiment Scenarios}
For our experiments, we considered 3 specific scenarios, as shown in Fig.~\ref{jpg5} and discussed in the following.
\subsubsection{Scenario 1}
Initially, we consider a simple environment consisting of three eNodeBs, each of which is connected to an SDN switch which is part of the backhaul network. Furthermore, an SDN switch implementing the serving gateway functionality is connected to each of the SDN switches, as shown in Fig.~\ref{jpg5} (Scenario 1). The SGW-SDN switch is connected to the primary gateway, which is connecting the transport backhaul to an external IP network.
\subsubsection{Scenario 2}
For Scenario 2, we assume a network of 7 eNodeBs, where each eNodeB is connected to an SDN switch in the backhaul network, as shown in Fig.~\ref{jpg5}.  The SDN switches connect the eNodeBs to the serving gateway switches via a capacity-limited link. The central eNodeB has a bottleneck link capacity that is half of its neighboring eNodeBs (as shown in Fig.~\ref{jpg5} Scenario 2).
\subsubsection{Scenario 3}
In Scenario 3, we use a similar 7 eNodeB network as in Scenario 2. However, the backhaul network is in this case replaced by a 100Mbps aggregation ring, and two 50 Mbps access rings, as shown in Fig.~\ref{jpg5} (Scenario 3). This scenario is used to evaluate our proposed scheme in the presence of BSs with different backhaul network bottleneck capacities, sharing links in a realistic network setting. 
\begin{table}[t!]
\caption{Simulation Parameters}\centering
\label{table:parameters}
\begin{tabular}{|c|c|}
\hline
\textbf{Simulation Parameters}   & \textbf{Value} \\
\hline
Number of macrocells & 3, 7 \\
\hline
Macro cell transmission power & 40 dBm \\
\hline
Macro cell radius & 250 m \\
\hline
Path loss model & Friis propagation path loss model \\
\hline
Scheduler  & CqaFfMacScheduler \\
\hline
Resource blocks (RBs) & 100 \\
\hline
Reuse factor & 1 \\
\hline
 Bandwidth & 20 MHz \\
\hline
 RSRQ threshold & 25 dB \\
\hline
 LB period ($\ell$) & 1 sec \\
\hline
$h_{\Delta}$ (Max-RSRQ handover) & 1 \\
\hline
\end{tabular}
\end{table}
%

\section{Results}
%
%
%
%
\subsection{Centralized Global vs. Distributed Local Optimization}
To compare the results of the local sub problems (solved at each eNodeB) with the globally obtained results, we used 12 randomly distributed users in a 3 eNodeB network (i.e. Scenario 1). Upon comparison, we noticed that the average DL throughput, when the optimization problem is solved locally at each eNodeB, is at least 90\% of the DL throughput results when the optimization is solved globally, for both uniform and non-uniform user distributions. Thus, in addition to having lower complexity and less overhead, the local and distributed load balancing approach also achieves significantly improved performance.

\subsection{Uniform Load Distribution}
For a uniform user placement, we uniformly randomly distributed users within 200 meter radius from the point central to the three eNodeBs of Scenario 1 of Fig.~\ref{jpg5}. 
\begin{itemize}
\item \textbf{GBR Users DL Performance.}
We performed experiments analyzing the DL data rate of GBR/QoS users by varying the number of users in the system. Fig.~\ref{jpg78}a shows the resulting average GBR user DL data rate achieved using different load distribution schemes by varying the network load. Note that as the numbr of users increases from 50 to 100, the achieved rate increases. However, it starts decreasing as the total number of users increases beyond 100. The initial increase in average rate is due to the topology used, i.e. as the number of users over the circle increases, users near the boundary of circular region (closer to eNodeBs) increases by a large proportion (due to bigger coverage area) and thus the average DL rate of GBR users increases. 
%
However, once the number of users reaches 150 (and beyond) the backhaul network is starting to become the bottleneck, and as a result, the DL rate of end users drops down to around 150 kbps.

Our proposed LB scheme results in better user DL rate performance compared to the other approaches. However, as the network load increases, the net benefit decreases slowly, due to the subsequent decrease of user efficiency in obtaining the desired rate metric. 
In summary, with a uniform user distribution, our proposed LB algorithm shows around 5\% improvement in DL rate, averaged over the considered range of network load (number of users).

\begin{figure}[t!]
\begin{center}
\resizebox{4.0 in}{!}{\includegraphics{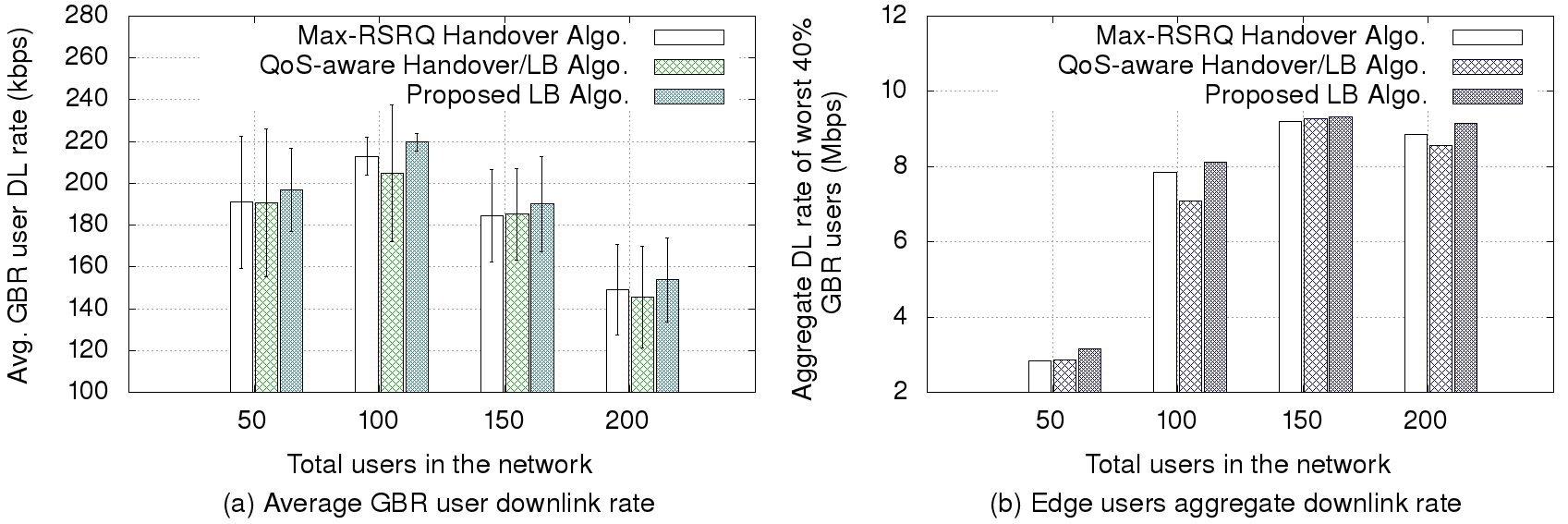}}
\caption{\em  Uniform users distribution in Scenario 1: Average of overall GBR users downlink rate and Edge users downlink rate}
\label{jpg78}
\end{center}
\end{figure}

\item \textbf{Edge User Network Performance.}
Here, we evaluate the aggregate DL rate of the lowest 40\% of the total GBR users (designated as the edge users) with respect to different network loads. As shown in Fig.~\ref{jpg78}b using our proposed algorithm, the edge user DL rates increase considerably compared to other approaches, especially when the network is congested. On average, our algorithm achieved an improvement in rate by edge users of around 5\%, compared to the Max-RSRQ based LB algorithm, and 8\% when compared to the QoS-aware handover algorithm.
\begin{figure}[b!]
\begin{center}
\resizebox{3.5 in}{!}{\includegraphics{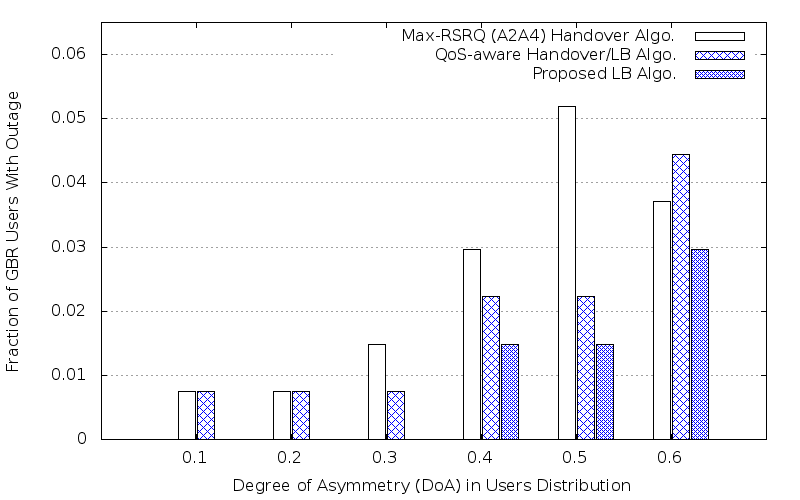}}
\caption{\em  Scenario 1 : Comparison of different handover algorithms with increasing degree of asymmetry (DoA) with  a total of 150 users in the network.}
\label{jpg9}
\end{center}
\end{figure}
\begin{figure*}[t!]
\includegraphics[width=\textwidth,height=4.0cm]{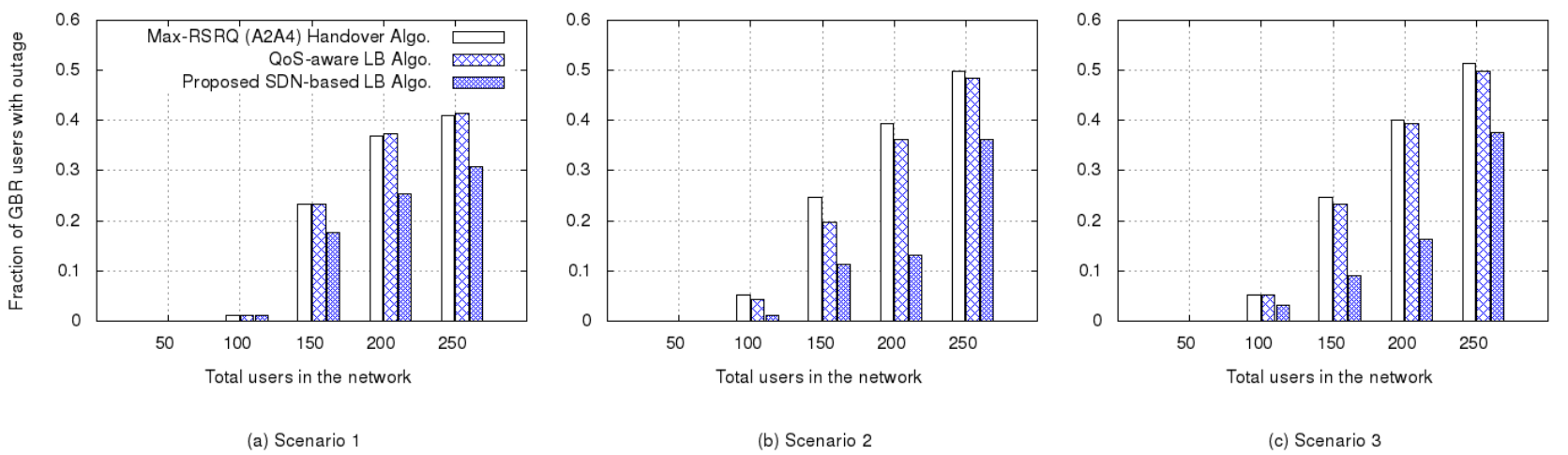}
\caption{\em Asymmetric User Distribution (Scenario 3): User outage in different scenarios }
\label{jpg10}
\end{figure*}
\begin{figure*}[t!]
\includegraphics[width=13.0cm,height=5.5cm]{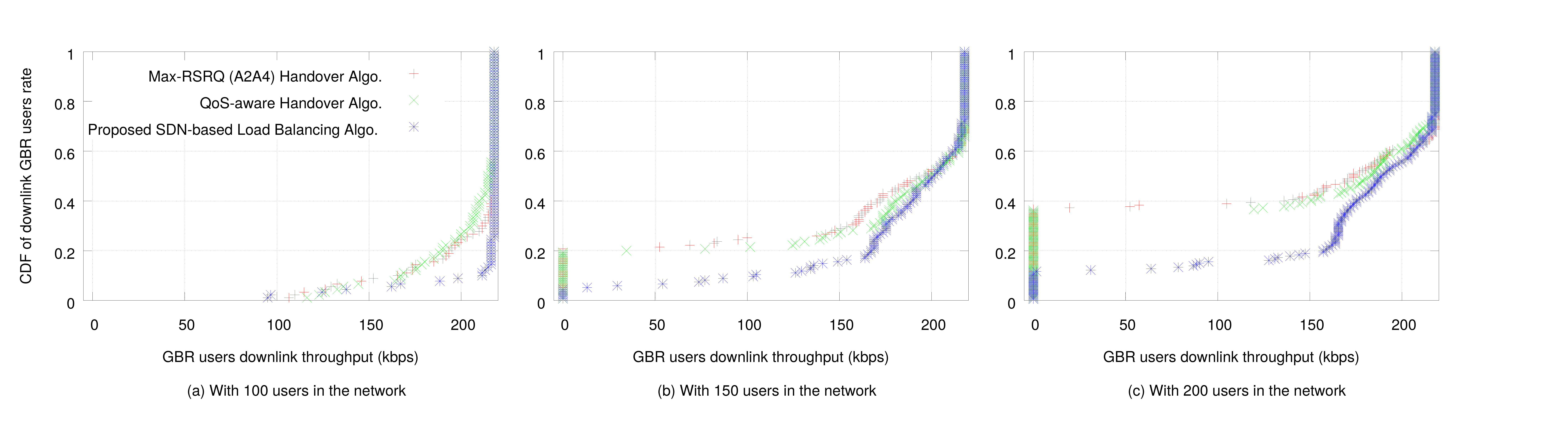}
\caption{\em Asymmetric User Distribution with 100, 150, and 200 users in the network: (a-c) Scenario 3}
\label{jpg11}
\end{figure*}
\item \textbf{Impact of Asymmetry.} 
In order to gradually shift from symmetric to asymmetric user distribution, we define the degree of asymmetry (DoA) parameter as the percentage of users over the loaded cell, compared to unloaded neighboring cells in the network. 
We study the resulting impact of asymmetry on the service deliver to users', i.e. service disruption or outage. In our case, the outage is defined as the fraction of GBR users experiencing a DL rate lower than 90 kbps~\cite{reference032} (desired rate for voice over LTE users). This can occur either due to limited backhaul or access network capacity, or due to bad channel quality. The results in Fig.~\ref{jpg9} compare the QoS rate performance obtained by increasing the DoA parameter over the coverage region for the three handover algorithms with a total of 150 users. 
For a small value of DoA, only the Max-RSRQ scheme has some users that cannot achieve the desired rate. As the DoA grows, the users rate outage increases in all three schemes. However, using our proposed LB scheme, most users are able to attain their desired rate performance. Also note a sharp increase in the fraction of users facing outage with the QoS-aware LB scheme as the DoA transitions from 0.5 to 0.6. The reason is that, when one cell gets increasingly overloaded, the QoS-aware LB approach associates some users to far away eNodeBs, resulting in a lower rate. 

In summary, our proposed LB mechanism achieves a 30\% improvement in terms of users achieving an acceptable rate, compared to the QoS-aware handover algorithm, and 80\% improvement compared to the Max-RSRQ algorithm. 
This is due to the fact that our proposed handover mechanism uses an improved objective function, and its ability to consider both the access and backhaul network when making handover decisions.

\end{itemize}

\subsection{Non-Uniform Load Distribution}
This section covers the results of our experiments carried out with an asymmetric user distribution over the coverage area. Here, one cell is more highly loaded compared to the neighboring eNodeBs. Each eNodeB coverage area is divided into an inner circle (of radius 170m) and an outer ring (radius from 170m to 250m) regions. We assume 50\% of all network users are uniformly randomly positioned over the outer 80m region of overloaded~\footnote{e.g. eNodeB 1 in Scenario 1, and eNodeB 4 in Scenario 2 and 3} cell. A further 30\% of users are uniformly distributed over all eNodeBs, while the remaining 20\% of the users are placed within the inner region of the overloaded cell.  
\begin{itemize}

\item \textbf{QoS Satisfaction} 
Fig.~\ref{jpg10}a-c shows the proportion of users that are unable to satisfy the desired rate requirement, either due to the weak signal strength or because of limited network resources. The results show that users unable to achieve their rates increases in different scenarios as the network gets more congested, while our proposed LB algorithms shows a minimal outage. Further note that our solution gives a higher gain in Scenarios 2 and 3, where there is more interference due to neighbouring cells and backhaul network congestion.
\item \textbf{GBR Users DL Performance.}
The results for the CDF of GBR users DL rates with various number of users in the network are shown for the third scenario in Figs.~\ref{jpg11}a to Fig.~\ref{jpg11}c. It can be seen that with our proposed LB algorithm, most users get considerably better DL rates. The result shows that with increasing users density, the gain achieved in the DL GBR user rate increases with the proposed LB approach. Similarly, the average GBR user DL rates results with respect to different number of users are shown in Fig.~\ref{jpg1112}a, Fig.~\ref{jpg1112}c, and Fig.~\ref{jpg1112}e. Compared to other approaches, the proposed load distribution method shows a promising performance, especially when the network traffic load is high. On average in Scenario 1, Scenario 2, and Scenario 3, our proposed algorithm shows (3\%, 4.7\%), (5\%, 7\%), and (12\%, 10.15\%) better GBR user DL rates, compared to the Max-RSRQ (A2A4) and QoS-aware handover algorithms, respectively.
\begin{figure}[t!]
\begin{center} 
\resizebox{3.6 in}{!}{\includegraphics{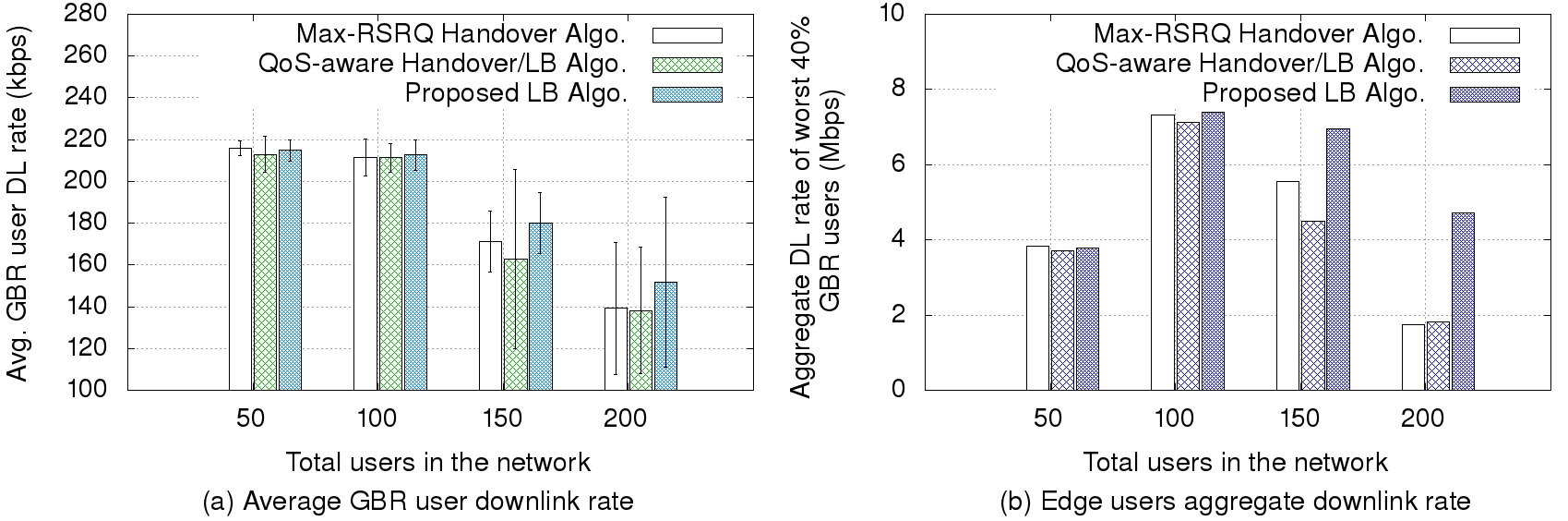}}\\
\resizebox{3.6 in}{!}{\includegraphics{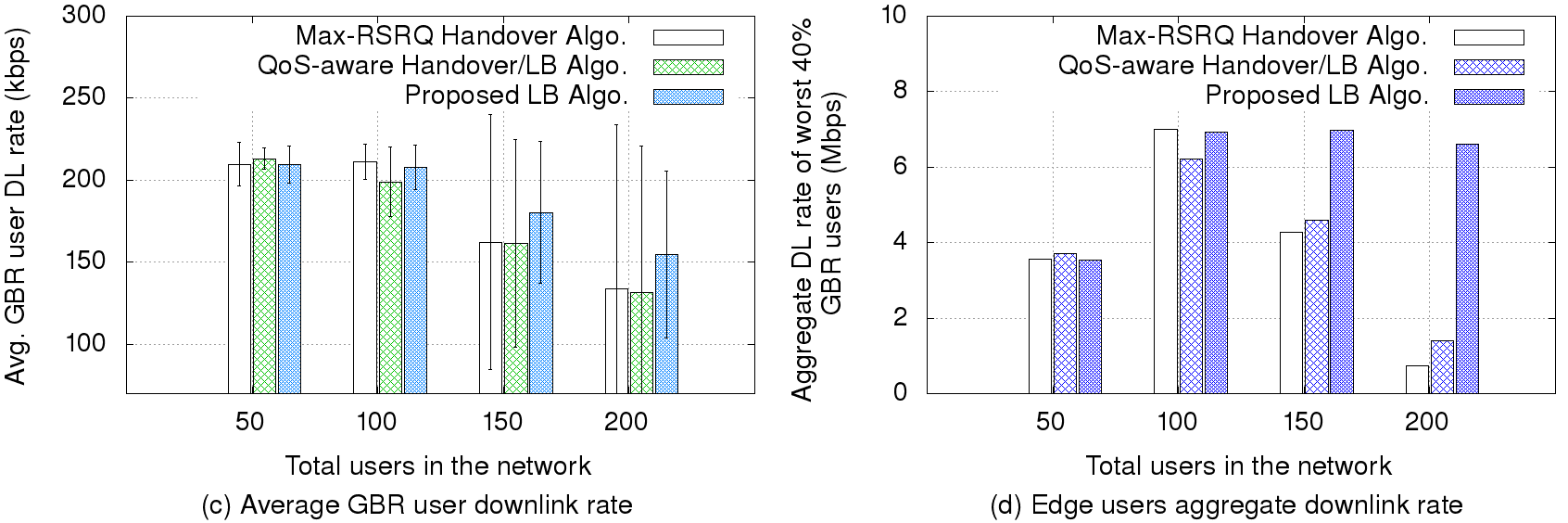}}\\
\resizebox{3.6 in}{!}{\includegraphics{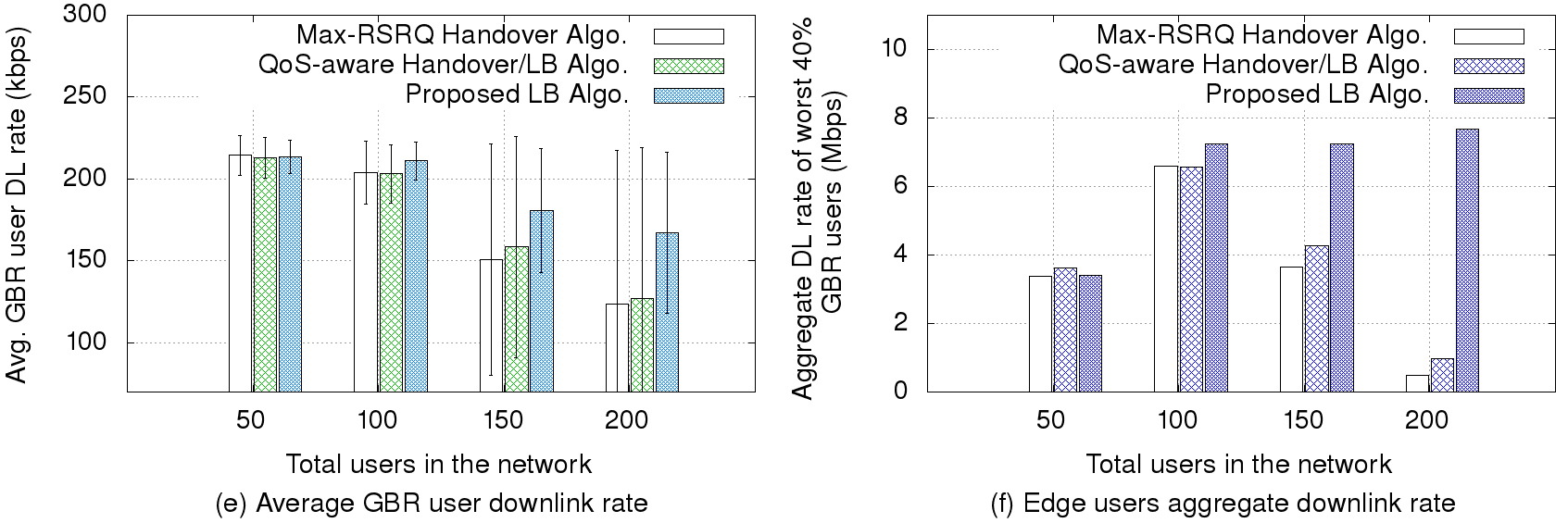}}
\caption{\em Asymmetric user distribution: (a,~b) Scenario 1, (c,~d) Scenario 2, (e,~f) Scenario 3.}
\label{jpg1112}
\end{center}
\end{figure}
\begin{figure*}[t!]
\includegraphics[width=13.0cm,height=5.5cm]{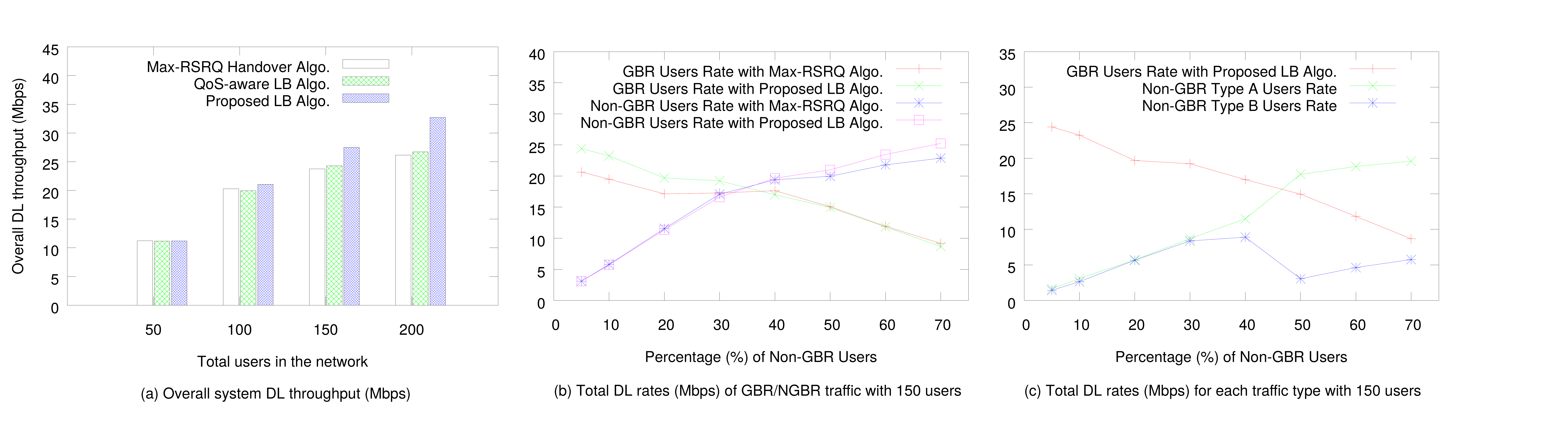}
\caption{\em Asymmetric user distribution, Scenario 3: (a) Overall System DL throughput (Mbps), (b) Total DL rates of GBR/NGBR users (Mbps) with 150 users, (c) Total DL rate (Mbps) of different users types (Type A: HTTP NGBR users, Type B: FTP NGBR users) with 150 users.}
\label{jpg1113}
\end{figure*}
\item \textbf{Edge User Network Performance.}
From the average DL rates results with different number of users (as shown in Fig.~9c) we observed that in the worst case (i.e. Max-RSRQ handover algo.) $\approx$ 40\% of total GBR users experience DL rates even lower than 90 kbps. Hence, we designate these users as boundary (edge) users. 

We further examine the achieved edge user aggregate DL rate obtained with a asymmetric users distribution in different scenarios, as shown in Fig.~\ref{jpg1112}b, Fig.~\ref{jpg1112}d, and Fig.~\ref{jpg1112}f. In all scenarios, the achieved performance of edge users using our proposed scheme is considerably better compared to the other load distribution approaches. Especially when the user load is high, and with more eNBs in a realistic backhaul setting, our approach achieves significantly higher gains. In different deployments (Scenario 1, 2, and 3), our algorithms shows an improvement of at least 12\%, 26\%, and 32\% respectively, in the average edge GBR user throughput.
\item \textbf{DL System Throughput.}
Fig.~\ref{jpg1113}a shows that the overall attained system throughput using our proposed LB approach is significantly higher compared to other approaches, as the network gets more highly loaded (e.g. approx. 32.7 Mbps when there are around 200 users in the network). Note that a greater fraction of throughput gain is achieved due to the GBR (/QoS) users, as the proposed approach uses both the access and backhaul network information to make a better handover decision as, shown in Fig.~\ref{jpg1113}a. Due to this, in high load condition (e.g. with 200 users), the proposed schemes prevents throughput degradation due to network congestion, by offloading users from the overloaded cell in a backhaul-limited LTE network.
%

\item \textbf{Effect of users traffic type.} Here, we explore the achieved total DL throughput of GBR and NGBR users by adjusting the ratio of NGBR to GBR users in the network. As seen from Fig.~\ref{jpg1113}b, our proposed SDN-inspired LB mechanism improves GBR and NGBR user total DL throughput when there are more GBR or NGBR users in the network. This is because when the network gets congested, the proposed mechanism is better able to optimally choose the candidate cell for handover. Our SDN-inspired user association algorithm shows 18.24\% of overall increase in GBR user rate when the traffic is dominated by GBR users, and a 10.4\% increase in NGBR user DL throughput when the traffic is dominated by NGBR flows.
\item \textbf{Congestion aware NGBR metering.} Next we assume two different NGBR user application traffic types, i.e. HTTP and FTP. As HTTP traffic flow are of higher priority than FTP, we use the QoS weight-age factor ratio of NGBR HTTP traffic to NGBR FTP traffic as 5:1. In our setup, separate weighted aggregate meters get activated for different NGBR flow types upon the detection of backhaul network congestion. 
As can be seen in Fig.~\ref{jpg1113}c, when the percentage of NGBR flows goes beyond 40\%, an increase in the aggregate DL rate of HTTP NGBR flows can be seen, whereas the DL rate of FTP NGBR flows decreases due to link congestion. Hence in our proposed SDN-based solution, the HTTP NGBR users are less affected due to the congested backhaul link, and achieve better congestion control with less overhead as compared to~\cite{reference002}.
\end{itemize}

\section{Further Considerations for SDN-enabled LTE Networks}
LTE 3GPP adopts a hard handover mechanism in which the overall handover process takes a maximum delay of 400ms-500ms~\cite{reference033, reference034}. This means, once the connection attach/de-attach procedure has completed, as illustrated in Fig.~\ref{jpg3} (procedure 6), the backhaul path switching should be completed within at most 200ms. The handover procedure in SDN-enabled LTE networks requires the controller to update the flow tables on SDN switches by inserting new, and/or deleting and modifying existing forwarding rules.
In contrast to networks with legacy devices, the SDN-based LTE backhaul can result in a significant reduction in the reconfiguration latency~\cite{reference035}. Since our evaluations are based on ns-3 simulations using a software switch, there can be some discrepancies to the path switching delays achieved with hardware SDN switches. However, we note that compared to traditional LTE networks, the handover latency in future SDN-LTE networks is relatively small due to its distributed architecture~\footnote{since multiple controllers will be distributed across the network infrastructure with each controller located close enough to the BS resulting in even lower latency} and the fewer required signaling messages~\cite{reference005,reference006}. 

For real SDN networks, the study in~\cite{reference031,reference036} shows that the latency during each of the above operations depends upon the vendor specific hardware switch, types of message processed, number of already installed rules and their priority order, as well as the CPU load at the switch. The work concludes that the overall path switching delay can scale from a few milliseconds (e.g. 5 ms) in case of an Intel hardware switch (due to efficient TCAM organization scheme) to a hundred milliseconds in case of a BCM-1.0 switch. 

Recently, different works have studied the problem of high flow insertion/update latency in SDN based networks. Two main factors contributing to this delay are an inefficient switch TCAM architecture, and schemes that check for black holes, loops, and link congestion during flow update. To tackle the first case,~\cite{reference037} suggests a modified TCAM control architecture that significantly reduces the overall flow update latency to below 5ms, typically suited for voice over LTE flows, while incurring a nominal increase in overhead of less than 5\%. For the second case, different recent works (e.g.~\cite{reference038, reference039}) studied the instant flow reconfiguration problem in SDN inspired networks. An effective combination of central and distributed schemes in~\cite{reference039} reduces the flow update latency to about 45\%. Similarly, ~\cite{reference040} suggests a four phases strategy to achieve consistent flow updates, while removing all deadlocks. In line with these advancements in SDN switch operation, we expect that future switch operation will further minimize the latency and achieve robust handover following the strict requirements of 5G (and beyond) networks. 
Finally, we conclude that the path switching delay ($\approx$ < 5ms) discussed above is relatively small to realize handover especially when an advanced SDN switch is in operation. 

In case of a rule insertion/modification failure, the OpenFlow switch v1.5.1 specifications~\cite{reference041} define an OFPT-ERROR-MSG structure through which a switch can notify the controller of such an error, e.g. that the TCAM memory is full. This allows the controller to take respective measures e.g. rechecking the flow-mod message, choosing an alternate route, etc. Note that in our experiments, due to the use of a software switch, we did not face any such issues. 

\section{Conclusion}
Our SDN-LTE network environment in ns-3 enables us to study and evaluate new scenarios to make future mobile networks simpler and more manageable. To make efficient handover decision in future SDN-enabled LTE networks, our work uses both the load information of the backhaul network as well as the access networks, together with consideration of fairness. Our findings show that with an asymmetric user distribution among neighbouring cells, our proposed load balancing and handover algorithm achieves significantly higher performance than state-of-the-art approaches. Also, our experimental evaluations also show a significant performance improvement for edge users. In future work, we are planning to extend this approach and more broadly explore new traffic management approaches based on a SDN-LTE integrated network.

\end{document}